\documentclass[3p]{elsarticle}

\usepackage{moreverb} 

\usepackage[table]{xcolor}
\usepackage{hyperref}
\usepackage{lineno}
\modulolinenumbers[5]
\usepackage{pgf}
\usepackage{tikz}
\usepackage{enumitem}

\usepackage{chngcntr}

\usepackage{float}
\usepackage{subcaption}
\usepackage{siunitx}

\usepackage{amsmath} 
\usepackage{amssymb}
\usepackage[short]{optidef}
\usepackage{bm}
\usepackage{bbm}
\DeclareMathOperator{\E}{E}

\usepackage{algorithmicx}
\usepackage{booktabs}
\usepackage[noend]{algpseudocode}

\usepackage[title]{appendix}

\usepackage{algorithm}

\usepackage{tabularx}
\usepackage{multirow}

\usepackage{collcell}

\def\colorModel{hsb} 

\newcommand\ColCell[1]{
	\pgfmathparse{#1<380?1:0}  
	\ifnum\pgfmathresult=0\relax\color{white}\fi
	\pgfmathsetmacro\compA{116/360}      
	\pgfmathsetmacro\compB{#1/700} 
	\pgfmathsetmacro\compC{1-#1/1000}      
	\edef\x{\noexpand\centering\noexpand\cellcolor[\colorModel]{\compA,\compB,\compC}}\x #1
} 
\newcolumntype{E}{>{\collectcell\ColCell}r<{\endcollectcell}}  

\usepackage{eurosym}
\algnewcommand{\Initialize}[1]{%
	\State \textbf{Initialize:}
	\Statex \hspace*{\algorithmicindent}\parbox[t]{.8\linewidth}{\raggedright #1}
}

\DeclareSIUnit{\EUR}{\text{\euro}}

\usetikzlibrary{arrows.meta}

\makeatletter
\g@addto@macro\UrlBreaks
{%
	\do\a\do\b\do\c\do\d\do\e\do\f\do\g\do\h\do\i\do\j%
	\do\k\do\l\do\m\do\n\do\o\do\p\do\q\do\r\do\s\do\t%
	\do\u\do\v\do\w\do\x\do\y\do\z\do\&\do\1\do\2\do\3%
	\do\4\do\5\do\6\do\7\do\8\do\9\do\0\do\/\do\.%
}
\g@addto@macro\UrlSpecials
{%
	\do\/{\mbox{\UrlFont/}\hskip 0pt plus 10pt}%
}

\journal{Applied Energy}

\biboptions{sort&compress}

\begin{document}

\begin{frontmatter}

\title{Techno-Economic Analysis and Optimal Control of Battery Storage for Frequency Control Services, Applied to the German Market}


\author[restore,kul,energyville]{Jonas Engels\corref{mycorrespondingauthor}}
\cortext[mycorrespondingauthor]{Corresponding author}
\ead{jonas.engels@restore.energy}

\author[restore]{Bert Claessens}

\author[kul,energyville]{Geert Deconinck}

\address[restore]{REstore NV/SA, 2600 Antwerp, Belgium}
\address[kul]{Department of Electrical Engineering, Div. Electa, KU Leuven, 3001 Leuven, Belgium}
\address[energyville]{EnergyVille, Thor Park 8310, 3600 Genk, Belgium}

\begin{abstract}
Optimal investment in battery energy storage systems, taking into account degradation, sizing and control, is crucial for the deployment of battery storage, of which providing frequency control is one of the major applications. 
In this paper, we present a holistic, data-driven framework to determine the optimal investment, size and controller of a battery storage system providing frequency control. We optimised the controller towards minimum degradation and electricity costs over its lifetime, while ensuring the delivery of frequency control services compliant with regulatory requirements. 
We adopted a detailed battery model, considering the dynamics and degradation when exposed to actual frequency data. Further, we used a stochastic optimisation objective while constraining the probability on unavailability to deliver the frequency control service. Through a thorough analysis, we were able to decrease the amount of data needed and thereby decrease the execution time while keeping the approximation error within limits. 
Using the proposed framework, we performed a techno-economic analysis of a battery providing 1 MW capacity in the German primary frequency control market. Results showed that a battery rated at 1.6 MW, 1.6 MWh has the highest net present value, yet this configuration is only profitable if costs are low enough or in case future frequency control prices do not decline too much. It transpires that calendar ageing drives battery degradation, whereas cycle ageing has less impact.

\end{abstract}


\end{frontmatter}


\section*{Acronyms}

\noindent
\begin{tabular}{ >{\bfseries}l l l l}
	BESS & Battery Energy Storage System & & \\
	BMS & Battery Management System & & \\
	CE & Continental Europe & & \\
	DoD & Depth of Discharge & & \\
	FCR & Frequency Containment Reserve & & \\
	HVAC & Heating, Ventilation, and Air Conditioning & & \\
	NMC & Nickel-Manganese-Cobalt & & \\
	RC & Resistance-Capacitance & & \\
	SAA & Sample Average Approximation & & \\
	SEI & Solid Electrolyte Interphase & & \\
	SoC & State of Charge & & \\
	TSO & Transmission System Operator & & \\
	UK & United Kingdom & & \\
	WAP & Weighted Averaged accepted bid Price & & \\
\end{tabular}

\vspace{-20pt}

\makebox[0pt][c]{
	\begin{tikzpicture}[remember picture, overlay]
	\node[anchor=south,yshift=50] at (current page.south) {\parbox{\dimexpr\textwidth-\fboxsep-\fboxrule\relax}{\copyright\,2019. This manuscript version is made available under the CC-BY-NC-ND 4.0 license.}};
	\end{tikzpicture}%
}

\section{Introduction}
Lithium-ion battery energy storage systems (BESSs) are being installed around the world at an increasing rate. An important application of BESSs is to provide frequency control or frequency regulation services. In multiple markets around the world, such as the market operated by PJM, in the UK or other energy markets in Europe, it is possible for third-party BESS operators to sell frequency control capacity to the transmission system operator (TSO). 

In future power systems, increased penetration of renewable generation and reduced inertia of large synchronous generators are expected to increase the need for fast frequency control reserves~\cite{GREENWOOD2017}. To mitigate this, battery energy storage systems are expected to play an important role as they are able to reduce volatility of the frequency of the grid, as has been shown in~\cite{LEE2019}, due to their rapid response time which cannot be matched by conventional generation assets.

Optimal investment, sizing and control are crucial for the deployment of BESSs to provide the required frequency control services.
However, performing a correct techno-economic analysis of a BESS is challenging, as there are a large number of non-linearities, parameters and uncertainties that need to be considered. 
Examples of these include the nonlinear dynamics and degradation of a battery cell, parameters of the control strategy (which is specific to each market) and uncertainties in the activation profile.
In this paper, we present an optimisation framework that considers these elements in detail, while still being able to compute in a reasonable amount of time. 
The framework determines the control strategy that minimises degradation while ensuring a delivery of the service compliant with the requirements of the TSO. Further, the framework allows to perform a techno-economic analysis, to calculate the investment case of a BESS over its lifetime and to determine its optimal size.



\subsection{Frequency Containment Reserve}\label{sec:intro_fcr}
In general, frequency control is divided into three distinct services: primary, secondary and tertiary frequency control. In this paper, we will focus on the primary frequency control service, or frequency containment reserve (FCR), as defined by ENTSO-E~\cite{ENTSO-E2013}, as it requires the fastest reaction time and least amount of energy content, making it very appropriate for a BESS. However, the framework presented in this paper can also be applied to secondary and tertiary frequency control services.

When providing FCR with an asset, the asset has to regulate its power output proportional to the deviation of the grid frequency from the nominal frequency (\SI{50}{Hz} in Europe). The maximum contracted reserve capacity should be activated when this frequency deviation reaches a predefined maximum value (\SI{200}{mHz} in the Continental Europe (CE) synchronous region) and within a predefined time interval (\SI{30}{s} in the CE region).


When having sold FCR capacity to the TSO in European FCR markets, one is required to deliver the service continuously during the contracted period. This is a problem for energy-constrained assets such as a BESS, because when a BESS is completely charged or discharged, it can no longer provide a symmetric service and faces penalties that are usually high (and can lead to exclusion from the market).
Therefore, an appropriate state-of-charge (SoC) controller or \emph{recharge controller} has to be in place, which maintains the SoC of the BESS within limits, ensuring the contracted FCR capacity is always available to be activated.

Note that this penalty mechanism as such does not exist in pay-for-performance frequency regulation markets in the USA, where one is paid according to a performance metric rather than penalised in case one does not deliver properly. Hence, the design requirements of the recharge controller in these pay-for-performance markets will also be different.



\subsection{Related Works and Contributions}
In the literature, quite a number of studies have been conducted on the use of a BESS for frequency control services, concentrating on different parts of the problem and using models with various degrees of detail. However, to the best of our knowledge, there is currently no work consolidating all elements with sufficient detail into one model.

The main focus of the work in \cite{HOLLINGER2016}, \cite{Stroe2017}, \cite{THIEN2017}, \cite{Zhang2016}, \cite{Cheng2016}, \cite{Xu2017} and \cite{LiFePO4Battery} is on the operational control strategy, including the recharge controller, of a BESS providing frequency control.
This control strategy should be designed carefully, as it has an important impact on the required energy content and on the lifetime of the BESS, as shown in~\cite{HOLLINGER2016}, \cite{Stroe2017} and~\cite{Kazemi2018}.
Specifically, in~\cite{Kazemi2018}, it was shown that it is important for the short-term operational control strategy to consider the long-term degradation for maximal revenues over the lifetime of the BESS, a conclusion that was also made in \cite{LEE2019}.

Rule-based recharge controllers, of which the parameters can be tuned, were proposed in~\cite{HOLLINGER2016}, \cite{Stroe2017}, \cite{THIEN2017}, \cite{MELO2019} and \cite{Lian2017} to provide FCR services to the German market.
More complex optimisation frameworks were proposed in~\cite{Zhang2016}, \cite{Cheng2016} and \cite{Xu2017}, albeit applied only to pay-for-performance frequency regulation markets. 
Dynamic programming was used in~\cite{Zhang2016} and \cite{Cheng2016}, but the results were operational control strategies that are computationally demanding and not feasible to calculate over the entire lifetime of the BESS, which is needed for investment analysis. 
In~\cite{Xu2017}, a control strategy that considers a more complex degradation model was optimised using a subgradient method. It was shown that a simple, rule-based controller can achieve a constant worst-case optimality gap with regard to a perfect-hindsight solution in pay-for-performance regulation markets.

A BESS is combined with a power-to-heat system to provide FCR services to the German market using a rule-based control strategy in~\cite{MELO2019}.
The combination of a BESS with a wind power plant to provide frequency control services was investigated in~\cite{JOHNSTON2015}, where they conducted an economic optimisation to determine the optimal size of the BESS.




A techno-economic analysis of a BESS performing frequency control with Li-ion battery cells was performed in~\cite{Fleer2018} for the German market, in~\cite{Lian2017} for the UK market and in~\cite{FARES2014} for the US market (Texas) but with a vanadium redox flow battery. Battery degradation was considered in both~\cite{Lian2017} and~\cite{Fleer2018}, but not in~\cite{FARES2014}. The operational control strategy was optimised in~\cite{Lian2017} via a grid search and in~\cite{FARES2014} using a nonlinear solver, but not in~\cite{Fleer2018}.

In \cite{HOLLINGER2016}, \cite{Stroe2017}, \cite{THIEN2017}, \cite{Zhang2016}, \cite{Cheng2016}, \cite{Xu2017}, \cite{MELO2019}, \cite{Lian2017}, \cite{JOHNSTON2015} and \cite{Fleer2018}, a simple, linear charge-counting battery model with constant efficiencies, energy and power capacities was used. In~\cite{Kazemi2018}, the efficiency losses were not considered.
Only in~\cite{LiFePO4Battery} and \cite{FARES2014} dynamic battery cell models were used, and it was argued that it is necessary to use accurate battery models when performing economic assessments.
This was confirmed in~\cite{BETZIN2018}, where they showed how the efficiency varies with the (dis)charging power when providing frequency control.

In each of these previous works, the focus was on a specific part of the problem: some works focussed on the design of the controller, but did not (or only to a limited extent) consider the dynamics or the degradation of the BESS or the stochastic nature of the FCR signal.
Other works focussed on the battery model or on the degradation of the BESS, but did not optimise the controller. 
In other works, a techno-economic analysis was performed, but without a dynamic battery model or an optimised controller. 
Hence, there is a clear need for a holistic approach, that allows conducting a complete techno-economic assessment of a BESS providing FCR using detailed models and an optimised FCR controller while considering the stochasticity in a correct way.

Therefore, in this work, we consolidated the results of previous works and appended to them the following contributions:
\begin{itemize}
	\item We present an all-encompassing framework for the investment analysis, sizing and control design of a BESS providing frequency control, featuring a dynamic BESS model, a semi-empirical degradation model and an optimised FCR controller that complies with current regulatory requirements.
	\item We propose a stochastic, data-driven optimisation algorithm that uses detailed historical frequency data and that allows constraining the probability on unavailability to a small value with high confidence. 
	\item We apply the framework to the German FCR market and analyse the results, which provides new insights into the economics and sizing of a BESS in this market.
\end{itemize}

The remainder of the paper is organised as follows: Section~\ref{sec:model_meth} elaborates the used models and the FCR controller. Section~\ref{sec:opt} presents the proposed optimisation algorithm. In Section~\ref{sec:results}, we discuss the application of the optimisation framework to the German FCR market and present the analysis of the results. Finally, the paper is concluded in Section~\ref{sec:conclusion}.

In the remainder of the paper, a bold symbol $\bm{x}$ denotes a vector containing the elements $x_i, i=1,\ldots n_x$, whereas a symbol with a bar $\overline{x}$ denotes the sample mean. We use $\lfloor x \rfloor $ to denote the floor function, which returns the greatest integer less than or equal to $x$; $\E[\cdot]$ the expected value operator; and $\mathbbm{1}\{x>x_0\}$, the indicator function, which returns $1$ if the value between brackets is true and 0 otherwise.



\section{BESS Model and FCR Controller}\label{sec:model_meth}

\begin{figure}
	\centering
	\hfill
	\begin{minipage}[b]{.45\textwidth}
	\centering
	\begin{tikzpicture}
	\node(rccell) [draw=black, text width=1.9cm, text centered] at (0,0) {Battery Cell Model};
	
	\node(degr) [draw=black, text width=1.9cm, text centered] at (0,1.2) {Degradation Model};
	
	\node(hvac) [draw=black] at (1.8,0.8) {HVAC};
	
	\node [inner sep=-0.05pt] at (3.0,0) (inv) {
		\begin{tikzpicture}[scale = 0.5]	
		\draw (0,0) rectangle (2,2);
		\draw (0,0) -- (2,2);
		\draw (0.2,1.8) -- (0.7,1.8);
		\draw (0.2,1.6) -- (0.7,1.6);
		\draw[domain=0:6.28, scale = 0.1] (1,0.2) plot (\x+12,{sin(\x r ) + 3});
		\end{tikzpicture} };
	
	\node [inner sep=-0.05pt] at (4.3,0) (grid){
		\begin{tikzpicture}
		\draw (0,0) circle [radius=0.3];
		\draw (0.3,0) circle [radius=0.3];
		\end{tikzpicture}};
	
	\node(contr) [draw=black, text width=1.9cm, text centered] at (3.0,2.4) {FCR Controller};
	
	\node(opt) [draw=black, text width=1.9cm, text centered] at (0,2.4) {Optimizer};
	
	\draw [->] (degr) -- (rccell);
	\draw [->] (contr) -- (inv);
	\draw [-{Implies[]}, double  distance=1.5pt] (opt) -- (contr);
	\draw [-{Implies[]}, double  distance=1.5pt] (0,1.8) -- (opt);
	
	\draw [ultra thick, -] (rccell) -- (inv);
	\draw [ultra thick] (inv) -- (grid);
	\draw [ultra thick] (hvac) -- (hvac |- inv);
	
	\draw[rounded corners, dashed] (-1.25,-0.65) rectangle (3.7,1.8);
	
	\node() at (1.4,-0.9) {BESS Model};
	\end{tikzpicture}
	\end{minipage}%
 	\hfill
	\begin{minipage}[b]{.45\textwidth}
	\centering
		\begin{tikzpicture}
	\draw[black, thick] (-2.5,0) -- (1.6,0);
	
	\draw[black, thick] (-2.5,0) -- (-2.5,0.5);
	\filldraw[black] (-2.65,0.5)rectangle(-2.35,0.6);
	\draw[black, ultra thick] (-2.9,0.7) -- (-2.1,0.7);
	\node [right] at (-2.1,0.6) {$V_{OC}(SoC)$};
	\draw[black, thick] (-2.5,0.7) -- (-2.5,1.7);
	
	\draw[black, thick] (-2.5,1.7) -- (-2.0,1.7);
	
	\draw[black, thick] (-2.0,1.3) -- (-2.0,2.1);  
	
	\draw[black, thick] (-2.0,2.1) -- (-1.7,2.1);
	\draw[black, thick] (-1.7,1.9)rectangle(-0.7,2.3);
	\node [above] at (-1.2,2.25) {$R_1$};
	\draw[black, thick] (-0.7,2.1) -- (-0.4,2.1);
	
	\draw[black, thick] (-0.4,1.3) -- (-0.4,2.1);
	
	\draw[black, thick] (-2.0,1.3) -- (-1.325,1.3);
	\draw[black,  very thick] (-1.325,1.0) -- (-1.325,1.6);
	\draw[black,  very thick] (-1.175,1.0) -- (-1.175,1.6);
	\node [above right] at (-1.2,1.3) {$C_1$};
	\draw[black, thick] (-1.175,1.3) -- (-0.4,1.3);
	
	\draw[black, thick] (-0.4,1.7) -- (0,1.7);
	
	\draw[black, thick] (0,1.5)rectangle(1,1.9);
	\node [above] at (0.5,1.85) {$R_0$};
	\draw[black, thick] (1,1.7) -- (1.6,1.7);
	\node [below] at (1.4,1.7) {+};
	\node [above] at (1.4,0) {--};
	\node at (1.4,0.85) {$V^{bat}$};
	\end{tikzpicture}
	\end{minipage}%
	\hfill
	\\[-7pt]
	\hfill
	\begin{minipage}[t]{0.45\textwidth}
 		\caption{Overview of the different models used in this study and their interaction.}
	 	\label{fig:Overview}	
 	\end{minipage}
 	\hfill
 	\begin{minipage}[t]{0.45\textwidth}
		\caption{First-order RC model of a battery cell.}
		\label{fig:RCmodel}
	 \end{minipage}
 	\hfill
\end{figure}

In this section, we elaborate the different parts of the BESS model and the FCR controller that we use in the optimisation. Figure~\ref{fig:Overview} gives an overview of all models used and their interaction. All parts in this figure will be discussed in this section one by one, except for the optimiser, which is discussed in Section~\ref{sec:opt}.

\subsection{Battery Cell Model}\label{sec:cellmodel}

Various types of battery cell models exist, each with its own level of detail and computational complexity. The most detailed cell models are the electrochemical models, such as the dualfoil model~\cite{fuller1994}, which try to capture in detail the various electrochemical processes that occur in the cells. These are typically the most accurate cell models, but require a large number of parameters and are computationally very demanding. 
Alternative analytical models, such as the kinetic battery model (KiBaM), are discussed in~\cite{jongerden2009}.

Lumped battery cell models or equivalent circuit battery cell models are often used because they require only a limited number of parameters, which provides a lower risk of overfitting compared to more complex models, while still attaining a good accuracy.
Among the equivalent circuit models, resistance-capacitance (RC) models of various orders are popular because of their simplicity and familiarity to the electrical engineering community. In~\cite{Hu2012}, Hu et al. made a comparison of 12 distinct equivalent circuit models for Li-ion battery cells.
They showed that, of the 12 equivalent circuit models, the first-order RC model, shown in Figure \ref{fig:RCmodel}, had the best performance, both on training and on unseen validation datasets. 
More specifically, as shown in \cite{LiFePO4Battery}, when providing frequency control, a purely resistive-based battery cell model already produces good results, with the main differences between the model and the measurements due to the absence of a capacitive element in the model.

As a compromise between accuracy and model complexity, we have used a first-order RC model in our BESS model, minimising the chances of overfitting. 
This RC model allows to capture the dynamics of the battery cells accurately, such as the variation in charging and discharging efficiencies with the current \cite{Li2015a}, which is neglected in simpler bi-linear battery models.

The parameters to be determined in this model are the ohmic resistances $R_0$ and $R_1$, the capacitance $C_1$ and the open circuit voltage $V_{OC}(SoC)$ as a function of the state of charge.
In this study, we modelled the Sanyo UR18650E battery cell~\cite{SanyoDatasheet}, a commercially available lithium-ion nickel-manganese-cobalt (NMC) cell with a graphite anode, which is one of the most common Li-ion cell chemistries in commercial grid storage battery systems. 
We used the results of Schmalstieg et al.~\cite{Ecker2014, Schmalstieg2014}, who created a detailed degradation model of this specific battery cell and provided enough information to determine the required parameters of the first-order RC model. 

\begin{figure}
	\centering
	\begin{subfigure}[t]{0.4\textwidth}
		\centering
		\includegraphics[width=\textwidth]{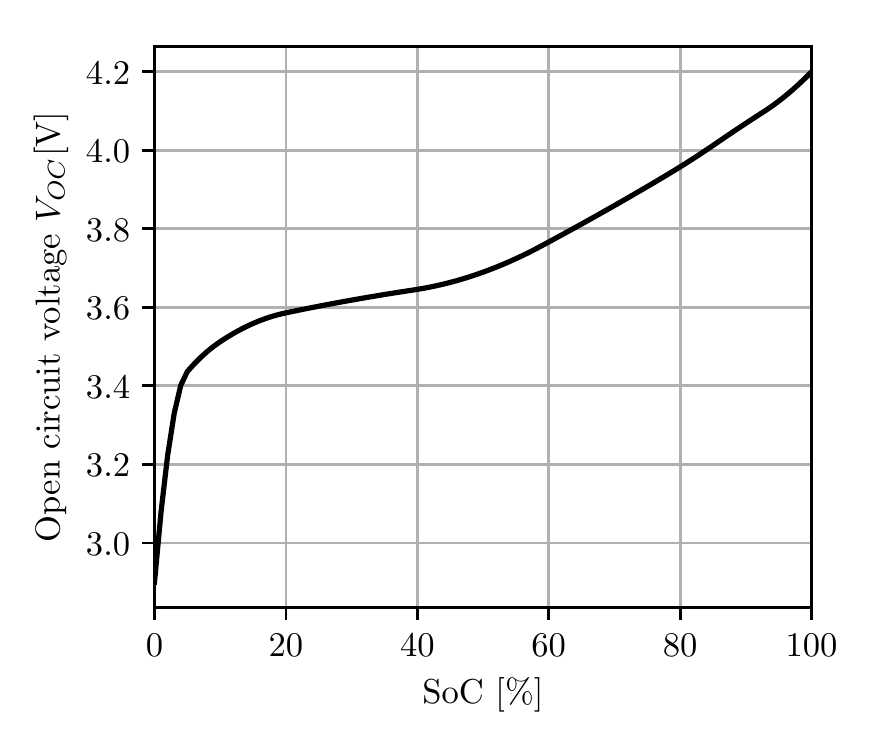}
		\caption{Open-circuit voltage curve $V_{OC}$}
		\label{fig:Voc}
	\end{subfigure}
	\hspace{2pc}
	\begin{subfigure}[t]{0.4\textwidth}
		\centering
		\includegraphics[width=\textwidth]{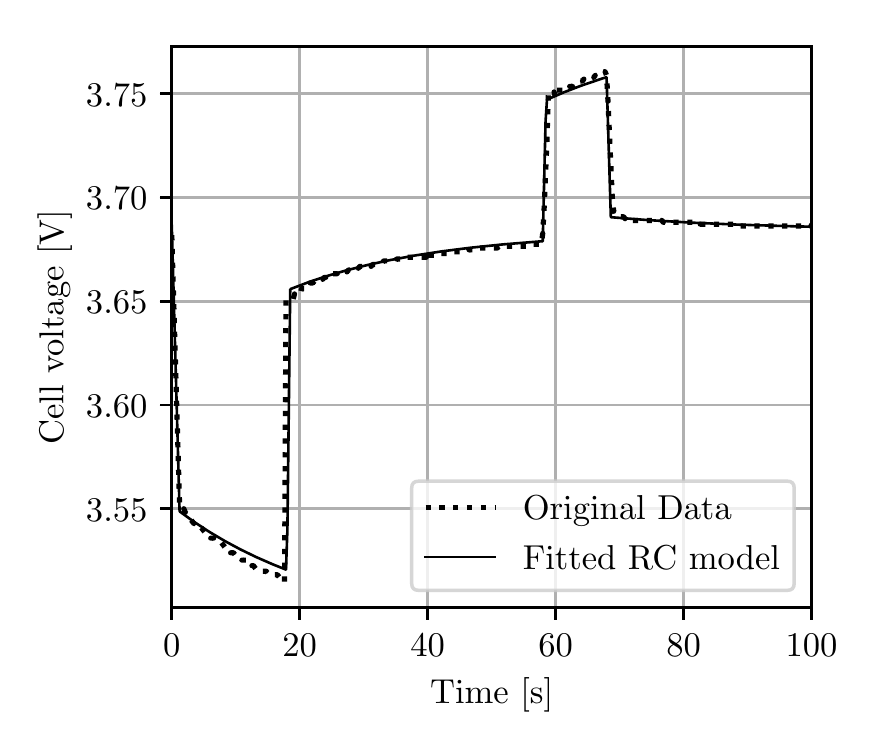}
		\caption{Pulse power test and fitted RC model response}
		\label{fig:FittedPulse}
	\end{subfigure}
	\caption{(a) Open-circuit voltage curve in the function of the SoC of the considered Li-ion NMC battery cell and (b) pulse power test and least-squares fit of the first-order RC model response, both from \cite{Ecker2014}.}
	\label{fig:BatteryCell}
\end{figure}

Figure~\ref{fig:Voc} shows the open-circuit voltage curve $V_{OC}(SoC)$ of the cell. We determined the remaining parameters from the battery cell voltage response to a pulse power test, shown in \ref{fig:FittedPulse}, using a least squares fit. The voltage response of the fitted RC model is also shown in~\ref{fig:FittedPulse}.
Table~\ref{tbl:Parameters} summarises the values of the fitted parameters of the RC model, together with some other key parameters of the battery cell.
The cut-off voltage when charging $V_{cutoff,charge}$ and discharging $V_{cutoff,discharge}$, which are the cell terminal voltages at which (dis)charging is stopped, was obtained from the Sanyo UR18650E datasheet~\cite{SanyoDatasheet}.
The heat capacity $C_p$ of the cell, needed for the heating, ventilation and air conditioning (HVAC) model, was retrieved from~\cite{SandiaThermal}.


%

\begin{table}[h]
	\centering
	\begin{tabular}{|l r | l r|}
	\toprule
		Parameter & Value & Parameter & Value  \\
		\midrule
		Nominal capacity $C$ & \SI{2.05}{Ah} & $V_{cutoff, charge} $& \SI{4.2}{V}\\
		Nominal resistance $R_0$ & \SI{0.0334}{\ohm} & $V_{cutoff, discharge}$ & \SI{2.75}{V}\\
		Nominal resistance $R_1$ & \SI{0.0114}{\ohm}& Nominal voltage $V_{nom}$ & \SI{3.6}{V}\\
		Nominal capacitance $C_1$ & \SI{1867.0}{F}& Heat capacity $C_p$ & \SI{40.05}{J/K}\\
		Coulombic efficiency $\eta_{coulomb}$ & \SI{99}{\%}& Rated energy capacity $E_{rated}$& \SI{7.38}{Wh}  \\
	\bottomrule
	\end{tabular}
	\caption{Parameters of the Sanyo UR18650E Li-ion NMC cell. The parameters of the first-order RC model ($R_0,R_1$ and $C_1$) were derived from the fit on the pulse power test profile shown in Figure \ref{fig:FittedPulse}.}
	\label{tbl:Parameters}
\end{table}

\subsection{Degradation Model}
Accurately quantifying the ageing or degradation of battery cells is important because degradation represents a capital loss of the battery investment costs. 
Unfortunately, degradation of battery cells is complex and not always well understood. Degradation originates from the interaction of various processes, complicating the identification of the root causes. 
Vetter et al.~\cite{Vetter2005a} gave a detailed qualitative overview of the various degradation processes in Li-ion batteries. Formation of the solid electrolyte interphase (SEI) on the anode is considered one of the most important sources of degradation. The SEI is a protective layer between the electrolyte and the anode, formed by decomposition of the electrolyte and accompanied by the irreversible consumption of lithium ions and a rise in impedance. 

Generally, battery degradation can be attributed to two factors: calendar ageing due to storage over time and cycle ageing due to repetitively charging and discharging of the battery cells. Barr\'{e} et al. \cite{Barre2013a} identified five different types of battery ageing models, ranging from detailed electrochemical models, such as extensions of the dualfoil model~\cite{Darling1998} and \cite{Ning2006}, to general statistical models. 

Empirical degradation models are often used due to their lower computational  complexity. These models result from experiments in which ageing of the cells is observed when these are exposed to various stress factors. For instance, a cell is stored at a certain SoC level or cycled with a certain depth of discharge (DoD) and the degradation is checked periodically.
A mathematical function, such as a polynomial or an exponential, is then used to describe the relationship between the applied stress factors and the observed degradation.

In their work~\cite{Ecker2014}, \cite{Schmalstieg2014}, Schmalstieg et al. described the ageing of the Sanyo UR18650E battery cell in detail. They described both capacity degradation and resistance growth when the cells were stored at various SoC levels and temperatures (calendar ageing), and when the cells were cycled around different SoC levels at various depths of discharge (cycle ageing). This results in an empirical model 
that correlates the SoC level and temperature during storage to the calendar capacity degradation and resistance growth with a $t^{0.75}$ time dependency, and the DoD and average SoC during cycling with the throughput $Q$ (in ampere hour) as follows:

%

\begin{subequations}
\begin{align} 
	C &= 1-\alpha_{cap}({SoC}^{cal}_{av},T) t^{0.75} - \beta_{cap}({SoC}^{cyc}_{av}, DoD) \sqrt{Q}, 
	\label{eq:CapDegr} \\
	R &= 1+\alpha_{res}({SoC}^{cal}_{av},T) t^{0.75} + \beta_{res}({SoC}^{cyc}_{av}, DoD) Q. \label{eq:ResDegr}
\end{align}
\label{eq:Degr}
\end{subequations}
Here, $\alpha_{cap}({SoC}^{cal}_{av},T)$ and $\alpha_{res}({SoC}^{cal}_{av},T)$ are the calendar ageing factors of capacity degradation and resistance growth, respectively, which are a function of the average state of charge during storage ${SoC}^{cal}_{av}$ and the temperature $T$ at which the cell is stored.
The cycle ageing factors $\beta_{cap}({SoC}^{cyc}_{av}, DoD)$ and $\beta_{res}({SoC}^{cyc}_{av}, DoD)$ on the other hand, are a function of the average state of charge ${SoC}^{cyc}_{av}$ during the cycle and the depth of discharge (in percent) $DoD$. 
The capacity degradation due to cycling has a square root dependency on the throughput $Q$, whereas the resistance growth shows a linear dependency on $Q$.

When performing frequency control the battery is cycled according to a stochastic profile rather than cycled repetitively with a constant depth of discharge. This makes the extraction of clearly defined cycles from the SoC profile not straightforward.
Therefore, we employed a \emph{rainflow} counting algorithm~\cite{cyclingStandard}, originating from material fatigue stress analysis to determine the cycles when materials are subject to an arbitrary load profile, but is also often used for cycle life assessment of batteries (e.g. in~\cite{Xu2017}, \cite{He2016} and \cite{Xu2016}). 
The rainflow counting algorithm takes as input the state of charge profile over time $\bm{SoC}\in\mathbb{R}^{n_t}$, with $n_t$ being the number of time steps. We adapted the original algorithm slightly to return, besides the DoD of a cycle, also the average state of charge of a cycle and the cumulative throughput $Q_{i_c}$ after each cycle $i_c = 1,\ldots,n_{cyc}$. The algorithm that implements the $Rainflow(\bm{SoC})$ function is detailed in \ref{sec:app_rain}:
\begin{equation}\label{eq:rainflow}
\bm{SoC}_{av}^{cyc}, \bm{DoD}, \bm{Q} = Rainflow(\bm{SoC}),
\end{equation}
where $\bm{SoC}_{av}^{cyc}, \bm{DoD}, \bm{Q} \in \mathbb{R}^{n_{cyc}}$, with $n_{cyc}$ being the number of cycles detected by the rainflow counting algorithm. 
We can calculate the capacity degradation after each cycle $i_c$ by integrating (\ref{eq:CapDegr}) over the throughput $Q$ as follows:
\begin{equation}\label{eq:cycle_degr}
	C_{i_c}^{cyc} =C_{i_c-1}^{cyc} - \int_{Q_{i_c-1}}^{Q_{i_c}}{\frac{\partial \beta_{cap}(SoC_{av,i_c}^{cyc},DoD_{i_c}) \sqrt{Q}}{\partial Q} d Q} = C_{i_c-1}^{cyc} - \beta_{cap}(SoC_{av,i_c}^{cyc},DoD_{i_c}) \cdot (\sqrt{Q_{i_c}} - \sqrt{Q_{i_c-1}}).
\end{equation}
An analogue reasoning is followed for the resistance growth due to cycling. To model calendar ageing, we determine the ${SoC}^{cal}_{av}$ in $\alpha_{cap}$ and $\alpha_{res}$ from (\ref{eq:Degr}) to be the average SoC of the entire profile $\bm{SoC}$.

In the remainder of the paper, we simulate the model for various operational years $k = 1,\ldots,n_k$ and use the index $k$ to denote the remaining capacity of the cell at the start of year $k$ by $C^k$ and the resistances by $R_0^k$ and $R_1^k$.

\subsection{From a Battery Cell Model to a BESS Model}
With the dynamic and degradation model of the battery cell determined, this section elaborates on how we used the cell model to simulate the behaviour of an entire BESS containing $n_{cells}$ cells. We did not model a battery management system (BMS), as we assumed that the BMS succeeds in balancing the cells in the battery pack perfectly and consumes a negligible amount of power. 
We also assumed that variations in cell characteristics are averaged out, allowing to simulate only one cell in detail, namely, the average cell, thereby drastically decreasing the simulation time. It is then straightforward to extrapolate the simulated power and SoC of the average cell proportionally to the required number of cells in the BESS.

Two other elements of the battery pack that cannot be neglected are the DC/AC inverter and the HVAC system.

\subsubsection{Inverter Model}
\begin{figure}
	\centering
	\includegraphics[width=0.4\textwidth]{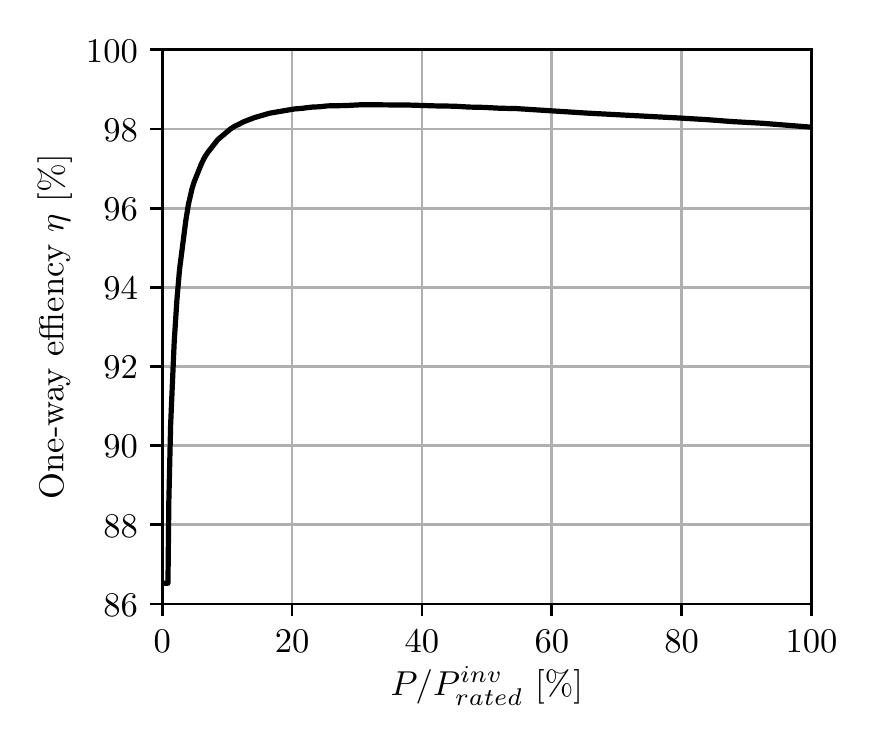}
	\caption{One-way efficiency of the inverter in function of its operating power relative to the rated power of the inverter $P_{rated}^{inv}$, based on the SMA STP60-10 model~\cite{SMA_STP60}.}
	\label{fig:InverterEff}
\end{figure}

Typically, the time constant of an inverter and its control system is an order smaller than the time constant needed for FCR. Therefore, we assumed that the inverter does not influence the dynamics of the BESS and can deliver any power required within one simulation time step, as long as this is possible within the capacity limits of the battery cell and the inverter.

The efficiency of an inverter is typically high, except at low power levels. 
Nevertheless, this can have considerable impact when performing frequency control, as the required power is often low and rarely reaches its maximum.
We modelled the efficiency of the inverter using the efficiency curve shown in Figure~\ref{fig:InverterEff}, taken from a commercial three-phase inverter (the SMA STP60-10~\cite{SMA_STP60}) which can be configured to deliver up to \SI{2.5}{MW} of power. We assumed the same efficiency curve for both consuming from and injecting into the grid. 
As the inverter is the gateway between the battery cells and the grid, the rated power of the inverter also determines the maximum power of the BESS: $P_{rated}^{inv} = P_{max}^{BESS}$.

\subsubsection{HVAC Model}
To determine the power consumption of the HVAC system, we employed a first-order thermal model of the battery cell, following \cite{Gatta2015}. From the first-order RC model of Figure \ref{fig:BatteryCell}, the Joule losses in the resistances $R_0^k$ and $R_1^k$ are dissipated as heat, thereby increasing the temperature of the cell $T$. This temperature is controlled by the HVAC system towards the reference temperature $T_{ref} = \SI{25}{\celsius}$. The thermal model of a system with $n_{cells}$ battery cells is governed by the following equation:
\begin{equation}
\label{eq:HVAC}
T_{t+1} = T_t +   \frac{ (R_0^k+R_1^k){I_t}^2 n_{cells} - COP\cdot P^{HVAC}_t }{ C_p n_{cells}} \Delta t,
\end{equation}
with $C_p$ being the heat capacity of the cell; $I_t$, the current in one cell at time step $t$; $COP$, the coefficient of performance, which we assumed to be $COP=2.5$; and $P^{HVAC}_t$, the instantaneous power of the HVAC system.
The Joule losses are equal to $ (R_0^k+R_1^k){I_t}^2 n_{cells}$ and $COP\cdot P^{HVAC}_t$ is the amount of heat removed by the HVAC system.
To prevent an unrealistically high HVAC power, we limited the power $P^{HVAC}_t$ to \SI{2}{\%} of the maximum power of the battery pack $P_{max}^{BESS}$.

\subsubsection{BESS Model}\label{sec:BessModel}
Putting together the cell model, the HVAC model and the inverter model, one obtains the following discretised model, which describes the dynamics of a BESS consisting of $n_{cells}$ battery cells required to deliver a certain power to the grid $P^{grid}_t$ at time step $t=1,\ldots,n_t$:
\begin{subequations}
\begin{align} 
P^{bat}_t &= \eta_{inv}(P^{grid}_t) \max(P^{grid}_t,0) + \frac{1}{\eta_{inv}(P^{grid}_t)}\min(P^{grid}_t,0) - P^{HVAC}_t, \label{eq:dyn_power} \\
I_t &= \frac{1}{2 R_0^k}\left(-V_{OC}(SoC_t) - V_t^{C_1} + \sqrt{\left(V_{OC}(SoC_t)+V^{C_1}_t\right)^2 + 4 R_0^k P^{bat}_t/n_{cells}}  \right) \label{eq:dyn_current}\\
V^{C_1}_{t+1} &= V^{C_1}_t e^{\Delta t/ (R_1^k C_1)} +  (1-e^{\Delta t/ (R_1^k C_1)}) R_1^k I_t, \label{eq:dyn_cap}\\
SoC_{t+1} &= SoC_t + \eta_{coulomb} \max(I_t,0) \frac{\Delta t}{C^k} + \frac{1}{\eta_{coulomb}} \min(I_t,0) \frac{\Delta t}{C^k}. \label{eq:dyn_soc}
\end{align} \label{eq:BESS_model}
\end{subequations}
The first equation calculates the required battery cell power $P^{bat}_t$ from the requested grid power $P^{grid}_t$ and the HVAC power $P^{HVAC}_t$, which results from (\ref{eq:HVAC}), while considering the inverter efficiency $\eta_{inv}(P^{grid}_t)$, which is dependent on $P^{grid}_t$ according to Figure \ref{fig:InverterEff}. Here, $P^{grid}_t>0$ when consuming from the grid and $P^{grid}_t<0$ when injecting into the grid and $P^{bat}_t > 0$ when charging and $P^{bat}_t < 0$ when discharging the battery cells.

Equation (\ref{eq:dyn_current}) translates the battery power divided by the number of cells $P^{bat}_t / n_{cells}$ into the battery cell current $I_t$, considering the voltage drop over the resistance $R_0^k$ and capacitor $V^{C_1}_t$. 
Equation (\ref{eq:dyn_cap}) represents the discretised dynamics of the parallel RC circuit $C_1, R_1^k$, whereas Equation (\ref{eq:dyn_soc}) represents the dynamics of the $SoC$ of the battery, with $C^k$ being the remaining capacity of the battery, $\eta_{coulomb}$ the coulombic efficiency and $\Delta t$ the duration of one time step. 
The BESS stops charging and discharging when the terminal voltage $V^{bat}_t = V_{OC}(SoC_t) + V^{C_1}_t +R_0^k I_t$ reaches $V_{cutoff, discharge}$ and $V_{cutoff, charge}$, respectively.

\subsubsection{Energy and Power Capacity of a BESS}\label{sec:E_P_cap_BESS}

\begin{figure}
	\centering
	\begin{subfigure}[t]{0.4\textwidth}
		\centering
		\includegraphics[width=\textwidth]{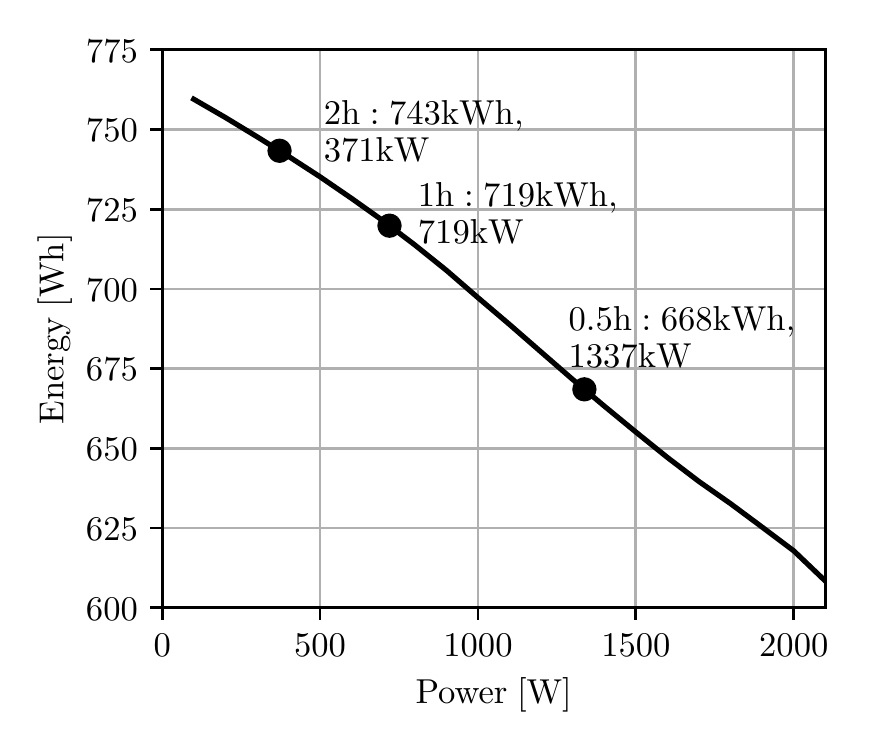}
		\caption{Available energy content  when charging and discharging at constant power.}
		\label{fig:energy_cPower}
	\end{subfigure}
	\hspace{2pc}
	\begin{subfigure}[t]{0.4\textwidth}
		\centering
		\includegraphics[width=\textwidth]{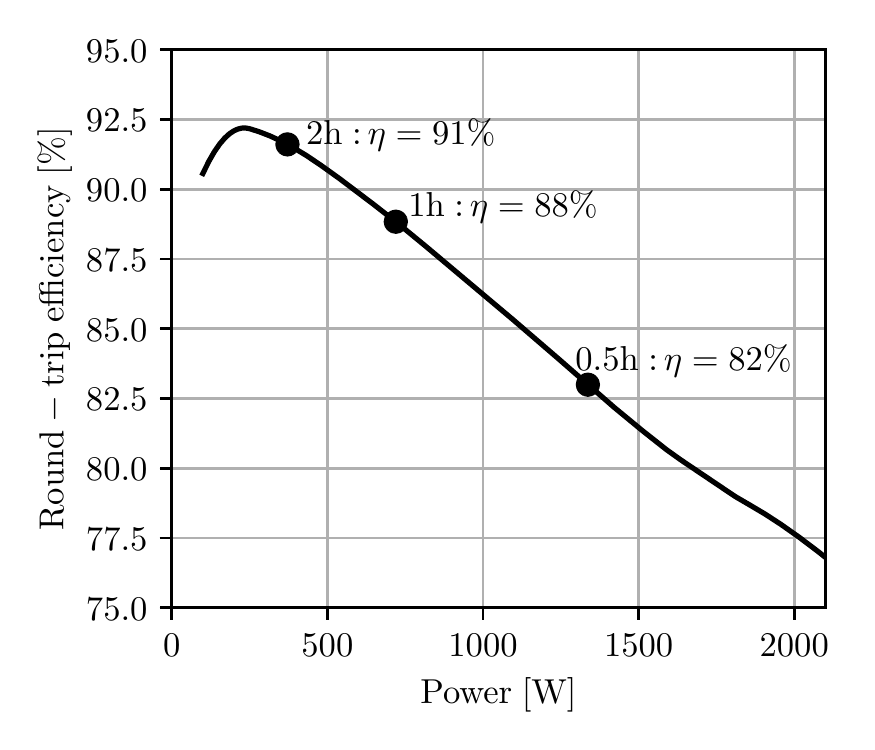}
		\caption{Round-trip efficiency $\eta$ when charging and discharging at constant power}
		\label{fig:eff_cPower}
	\end{subfigure}
	\caption{(a) Available energy and (b) round-trip efficiency $\eta$ when charging and discharging at constant power until the cut-off voltage is reached, using the BESS model with 100 battery cells, inverter rated at 2100 W and a 42 W HVAC system.}
	\label{fig:constant_power}
\end{figure}

The capacity of a battery cell is usually expressed in ampere hour (Ah), whereas the energy capacity of a commercial BESS is usually expressed in kilowatt hour (kWh). 
Although the energy content of the cell is rated at $\SI{7.35}{Wh}$, the actual energy that can be charged or discharged is dependent on the (dis)charging current. A higher current will induce greater losses in the resistive elements and thus provide less usable energy. Moreover, the voltage drop over the resistive elements will mean that the cutoff voltage $V_{cutoff, discharge}$ will be reached earlier and discharging will stop before the SoC reaches \SI{0}{\%}.
An analogue reasoning holds when charging the battery cell.



This effect is quantified in Figure \ref{fig:energy_cPower}, which shows the available energy capacity of the simulated BESS system containing $n_{cells}= 100$ battery cells, an inverter rated at $P_{rated}^{inv} = \SI{2100}{W}$ and a \SI{42}{W} HVAC system when charging at constant power until the terminal voltage $V^{bat}$ reaches $V_{cutoff, charge}$ and subsequently discharging at constant power until $V^{bat}$ reaches $V_{cutoff, discharge}$.
As can be seen in the figure, the available energy capacity of the BESS decreases with an increase in power, due to an increase in losses, and reaches the cut-off voltages earlier. 
Figure \ref{fig:eff_cPower} shows the round-trip efficiency of the same BESS in the function of the (dis)charging power, in line with the experimental results from~\cite{BETZIN2018}.
Because of an increase in resistive losses in the battery cells, the round-trip efficiency of the BESS decreases with an increase in power.
However, at low power, the efficiency of the BESS decreases as well. This decrease is due to the low efficiency of the inverter at low power rates (as shown in Figure \ref{fig:InverterEff}) rather than to efficiency losses in the battery cells themselves. At higher power rates, the efficiency of the inverter has less impact as it is rather high and nearly constant.

Finally, in battery cells, the rate-capacity effect \cite{Doyle1997} (also described by Peukert's law~\cite{peukert1897}) also limits the available capacity of the cell when discharged at higher currents. 
We did not explicitly model the rate-capacity effect, as it has been shown that it does not hold when operating the cell at variable currents \cite{Doerfell2006}, which is the case when performing frequency control services.

\subsection{FCR Controller}\label{sec:model_controller}
When providing frequency containment reserves, one has to adjust its power for FCR proportionally to the relative deviation of the frequency of the grid from the nominal frequency: $P_t^{FCR} = r\Delta f_t= r(f_t - f_{nom}) / \Delta f_{max}$, so that the contracted FCR capacity $r$ is reached at a maximum predefined frequency deviation $\Delta f_{max}$.

As explained in Section~\ref{sec:intro_fcr}, a recharge controller $\pi(SoC)$ that controls the SoC is necessary when participating in FCR markets with energy-constrained assets. 
In the literature, different versions of such recharge controllers have been proposed, ranging from simple rule-based controllers in \cite{HOLLINGER2016}, \cite{Stroe2017}, \cite{THIEN2017},\cite{Lian2017}, \cite{Oudalov2007} and \cite{Delfanti2014}, to moving average filters in \cite{Borsche2013}, \cite{Megel2013} and linear state-feedback controllers optimised using robust optimisation in~\cite{Engels2017}.

\begin{figure}
	\centering
	\hfill
	\begin{minipage}[b]{.47\textwidth}
		\centering
		\includegraphics[width=\textwidth]{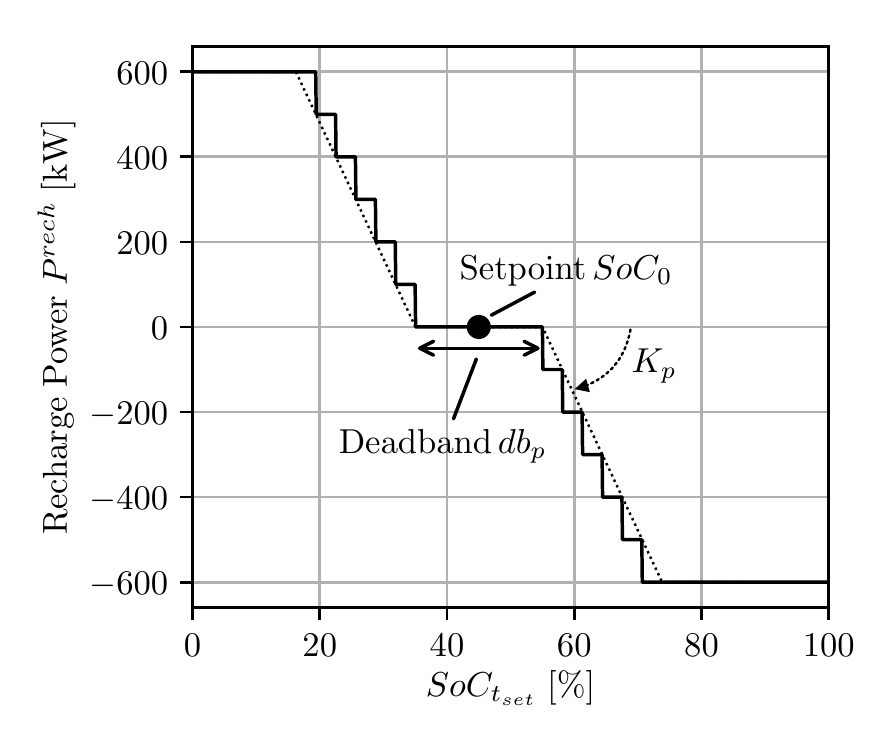}
	\end{minipage}%
	\hfill
	\begin{minipage}[b]{.47\textwidth}
		\centering
		\includegraphics[width=\textwidth]{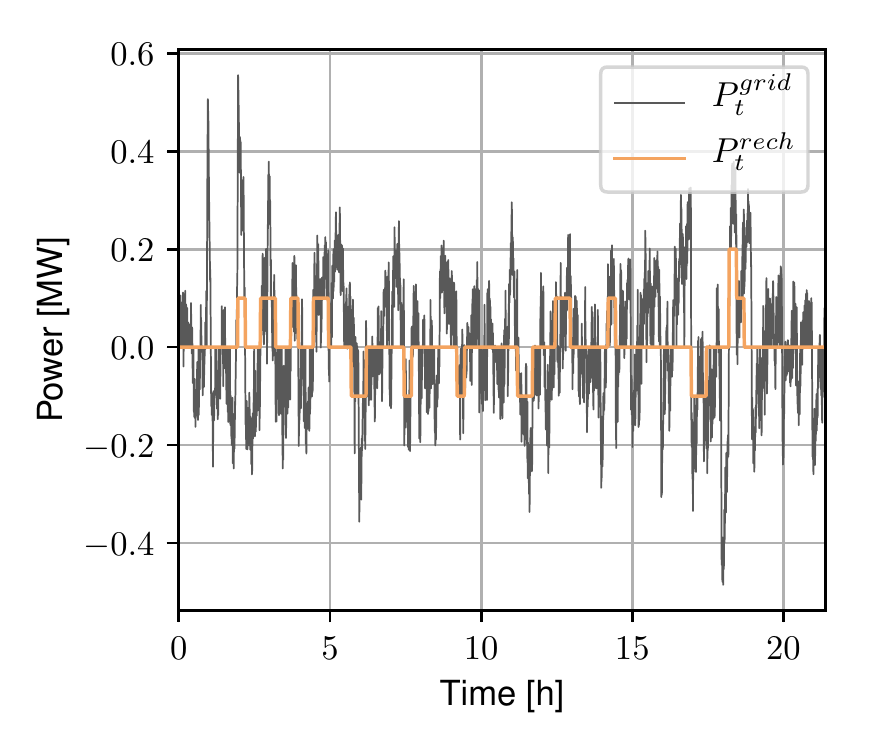}
	\end{minipage}%
	\hfill
	\\[-7pt]
	\hfill
	\begin{minipage}[t]{0.47\textwidth}
		\caption{Example of a possible recharge controller function $P^{rech}_t = f(SoC_{t_{set}})$ of a \SI{1.6}{MW}/\SI{1.6}{MWh} BESS delivering $r=\SI{1}{MW}$ of FCR capacity, with $K_p=2, SoC_0 = 0.45$ and $db_p = 0.2$. The broken line represents a linear P-controller, whereas the black line is the version discretised to multiples of \SI{100}{kW}, as implemented in this study.}
		\label{fig:Recharge_controller}	
	\end{minipage}
	\hfill
	\begin{minipage}[t]{0.47\textwidth}
		\caption{Example of the grid power $P^{grid}_t$ and recharge power $P^{rech}_t$ of a \SI{1.6}{MW}/\SI{1.6}{MWh} BESS delivering $r=\SI{1}{MW}$ of FCR capacity, with the recharge controller from Figure~\ref{fig:Recharge_controller}.}
		\label{fig:Power}
	\end{minipage}
	\hfill
\end{figure}

In this study, we adopted a simple, discretised P-controller $f(\cdot)$ with a deadband $db_p$ and a proportional gain $K_p$ that controls the SoC back to a setpoint $SoC_0$, shown in Figure \ref{fig:Recharge_controller}.
The output of the proportional error is discretised in steps of \SI{100}{kW}, kept constant for a time period $t_{recharge}$ and determined 
upfront with a lead time $t_{lead}$ to be compliant with the requirements of the German FCR market~(see Section~\ref{sec:GermanFCR_BESS}): $P^{rech}_t= f(SoC_{t_{set}})$, with $t_{set} = \lfloor t / t_{recharge} \rfloor t_{recharge} - t_{lead}$.
As the recharge power cannot be used as FCR capacity at the same time, the maximum recharge power $P^{rech}_{max}$ is limited to the maximum power of the battery minus the FCR capacity: $|P^{rech}_t|\leq P^{rech}_{max}\leq P_{max}^{BESS} - r$.


Besides specifically reserving recharge power, we also implemented overdelivery as a way to recharge the battery. When overdelivering, the BESS delivers more power than required (in absolute value). In our controller, we perform overdelivery only when this is beneficial to get the SoC back to the setpoint:

\begin{equation}
P^{od}_t = \begin{cases}
o_d  r \Delta f_t &\text{if $sign(SoC_t-SoC_0)=-sign(\Delta f_t)$}, \\
0 &\text{otherwise}, 
\end{cases}
\end{equation}
with $o_d$ being the percentage of overdelivery. 
The total power at the grid $P^{grid}_t$ is then the sum of the power for FCR $P_t^{FCR} = r\Delta f_t$, the recharging power $P^{rech}_t$ and the power for overdelivery $P^{od}_t$, for every time step~$t$ :
\begin{equation}\label{eq:FCR_controller}
P^{grid}_t = r \Delta f_t +P^{rech}_t + P^{od}_t,
\end{equation}
which is limited by the maximum power $|P^{grid}_t| \leq P_{max}^{BESS}$ of the BESS.
This controller can be seen as an extension of the rule-based controllers proposed in \cite{HOLLINGER2016}, \cite{Stroe2017}, \cite{THIEN2017}, \cite{Lian2017}, \cite{Oudalov2007} and \cite{Delfanti2014}, and as a special case of the ones in \cite{Borsche2013}, \cite{Megel2013} and \cite{Engels2017}.

Figure~\ref{fig:Power} shows an example of the grid power $P^{grid}_t$ of a battery delivering $r=\SI{1}{MW}$ of FCR capacity, according to (\ref{eq:FCR_controller}). The figure also shows the corresponding recharge power $P^{rech}_t$ according to the recharge controller from Figure~\ref{fig:Recharge_controller}.

%
%
%

\section{Optimisation Framework} \label{sec:opt}
The FCR controller of the BESS discussed in the previous section has four parameters (i.e., the deadband $db_p$, the setpoint $SoC_0$, the proportional gain $K_p$ and the amount of overdelivery $o_d$), which can be chosen independently. These parameters determine how the battery will be used, how fast it degrades and how much the electricity costs will be. For instance, increasing the deadband $db_p$ reduces the throughput but increases the width of the SoC distribution and, thus, the DoD of the cycles, whereas increasing the overdelivery parameter $o_d$ increases the throughput but reduces the probability on penalties due to unavailability of the BESS.  

To determine the value of these parameters, we defined the following optimisation problem, which maximises the revenues from providing $r$ FCR capacity taking into the electricity costs $c_{elec}^k$ and the degradation $\Delta C^k$, while constraining the probability on penalties $p^k$:
\begin{mini!}
{\bm{x} \in \mathcal{X}}{  -\E[c_{FCR}^k] r + \E[ c_{elec}^k(\bm{x}, \bm{\Delta f})] + \frac{\E[\Delta C^{k}(\bm{x}, \bm{\Delta f})]}{\SI{100}{\%}-\SI{80}{\%}} c_{cell} \label{eq:opt_problem_obj}}
{\label{eq:opt_problem}}{},
\addConstraint{\Pr\{ p^k(\bm{x}, \bm{\Delta f}) > 0 \} \leq \epsilon^{req}, \label{eq:opt_problem_constr}}
\end{mini!}
with $\bm{x} = (K_p, SoC_{0}, o_{d}, db_{p}) \in \mathcal{X}$ being the decision variables constrained to the admissible set $\mathcal{X} \subset \mathbb{R}^4$ and $\bm{\Delta f} = (\Delta f^0, \Delta f^1, \ldots, \Delta f^{n_{t,y}})$ being the stochastic frequency deviation of length $n_{t,y} = 365\times24\times3600/\Delta t$, covering one year.
As the frequency of the grid and, thus ,the required battery power are unknown upfront, a probabilistic approach is required.
The optimisation is to be executed each year $k = 1, \ldots, n_k$ the BESS is operational in FCR, allowing the adjustment of the controller parameters as the battery degrades, for instance, decreasing the deadband $db_p$ when less energy capacity is remaining.

The objective function is a compromise between three factors: the revenues from delivering FCR with a capacity $r$ at an expected price $\E[c_{FCR}^k]$, the electricity costs $c_{elec}^k$ and the degradation of the battery $\Delta C^{k}$. 
Whereas the first two are easily expressed in monetary value, we assign a value to the last one, which is equal to the cost of replacing the battery cells $c_{cell}$ times the incremental degradation $\Delta C^{k} = C^{k}-C^{k+1}$ from (\ref{eq:CapDegr}), divided by $(\SI{100}{\%}-\SI{80}{\%})$, as the battery's end of life is assumed to be reached when $C=\SI{80}{\%}$. 
We employ the expected value operator $\E[\cdot]$ over the electricity costs and the degradation as both are dependent on the actual frequency deviation profile $\bm{\Delta f}$, which is stochastic by nature.
The optimisation is constrained by a chance constraint (\ref{eq:opt_problem_constr}), forcing the probability of incurring penalties $p^k(\bm{x},\bm{\Delta f})$ to be less than or equal to $\epsilon^{req}$. 
The functions are indexed with $k$ to denote their dependence on the remaining battery capacity $C^k$ and resistances $R_0^k, R_1^k$ at the start of year $k$.

It is interesting to note that the optimisation problem (\ref{eq:opt_problem}) is part of the family of dynamic programming problems~\cite{bertsekas2005dynamic}. Indeed, (\ref{eq:opt_problem}) is actually a policy search over the policies parametrised by $\bm{x}$, with the last term of (\ref{eq:opt_problem_obj}) being a heuristic approximation of the value function $V(C^{k+1})$ of the next state $C^{k+1}$.

In the next subsections, we will elaborate first on how we approximate the expected value operators in the objective function, then on how we deal with the chance constraint on the penalty and, finally, on the global optimisation algorithm we employ to solve (\ref{eq:opt_problem}).



\subsection{Expected Value Approximation}
The objective function (\ref{eq:opt_problem_obj}) consists of the sum of three expected value operators. The first one, the expected FCR price $\E[c_{FCR}^k]$ during year $k$, is independent of the decision variables $\bm{x}$ and can thus be evaluated before the start of the optimisation routine.

The second and third expected value operators, however, do depend on the decision variables $\bm{x}$ and will thus have to be approximated when evaluated during optimisation. A general approach is to use a sample average approximation (SAA)~\cite{Shapiro2007} of the expected value by taking the empirical mean over independent and identically distributed (iid) samples of the stochastic variable, i.e. the frequency deviation profile $\bm{\Delta f}$. 
Let $\bm{\Delta f}_y \in \mathcal{Y}$ be the set of available frequency deviation samples of one year, with  $|\mathcal{Y}|=n_\mathcal{Y}$, then:
\begin{subequations}
\begin{align}
\E[ c_{elec}^k(\bm{x}, \bm{\Delta f})] & \approx \frac{1}{n_\mathcal{Y}} \sum_{\bm{\Delta f}_y\in\mathcal{Y}} c_{elec}^k(\bm{x}, \bm{\Delta f}_y), \label{eq:SAA_eleccost}\\
\E[\Delta C^{k}(\bm{x}, \bm{\Delta f})] & \approx \frac{1}{n_\mathcal{Y}} \sum_{\bm{\Delta f}_y\in\mathcal{Y}} \left( \Delta C^{cal,k}(\bm{x}, \bm{\Delta f}_y) +  \Delta C^{cyc,k}(\bm{x}, \bm{\Delta f}_y)\right). \label{eq:SAA_degr}
\end{align}\label{eq:SAA}
\end{subequations}
Following (\ref{eq:CapDegr}), the incremental degradation of the battery $\Delta C^k$ in (\ref{eq:SAA_degr}) has been written down directly as the sum of calendar degradation $\Delta C^{cal,k}$ and cycle degradation $\Delta C^{cyc,k}$.

\subsubsection{SAA Using Frequency Samples of One Day}
Problem (\ref{eq:opt_problem}) optimises one year of operation of the BESS and thus needs various frequency samples of one year $\bm{\Delta f}_y \in \mathbb{R}^{n_{t,y}}$ for an accurate SAA (\ref{eq:SAA}). As simulating multiple samples of one year of frequency data generally takes too long to employ during each step of an optimisation algorithm, we used $n_\mathcal{D}$ iid frequency samples of one day $\bm{\Delta f}_d \in \mathcal{D} \subset\mathbb{R}^{n_{t,d}}, n_{t,d} = n_{t,y}/365$ instead of one year in (\ref{eq:SAA}) during the optimisation. The electricity costs of one year in (\ref{eq:SAA_eleccost}) can then be retrieved by linear extrapolation of the electricity costs of one day: $c_{elec}^k(\bm{x}, \bm{\Delta f}_y) \approx 365  c_{elec}^k(\bm{x}, \bm{\Delta f}_d)$.

\begin{figure}
	\centering
	\includegraphics[width=0.45\textwidth]{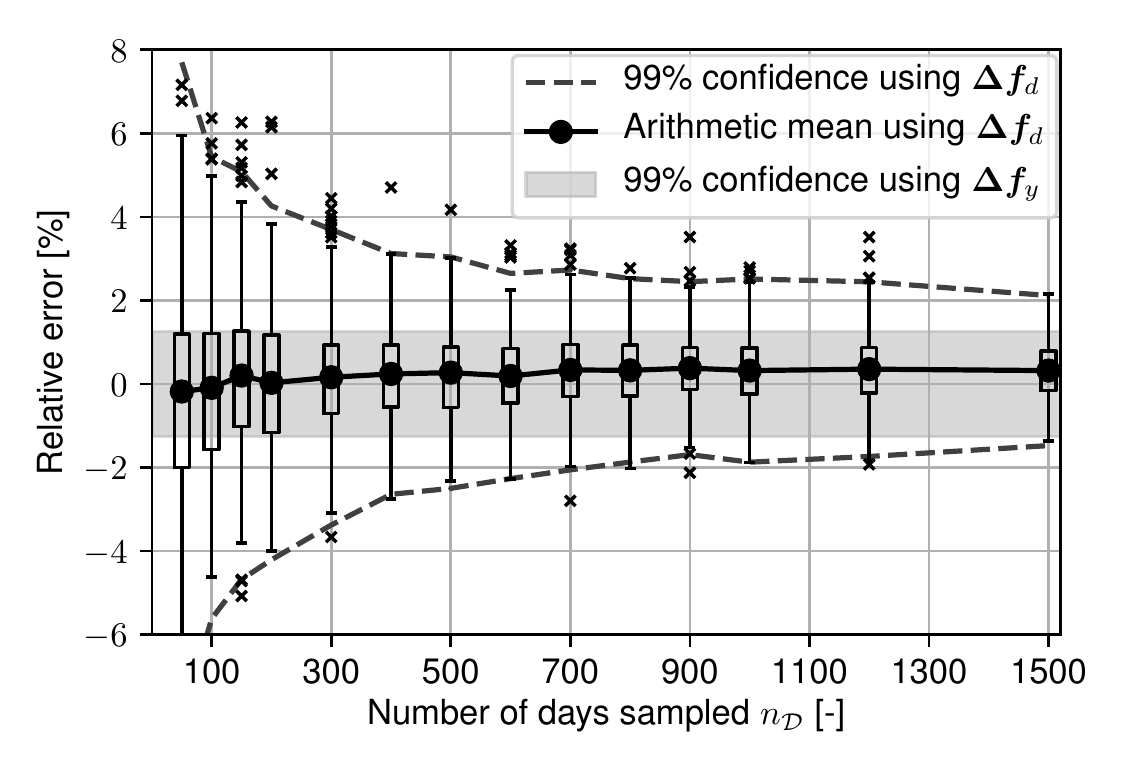}
	\caption{Boxplot of the relative error when approximating the SAA of  (\ref{eq:opt_problem_obj}) with frequency samples of one day $\bm{\Delta f}_d$ and using (\ref{eq:degr_days_appr}), compared to the SAA using frequency samples of one year $\bm{\Delta f}_y$. The \SI{99}{\%} confidence intervals of the error using samples of one day $\bm{\Delta f}_d$ and of the SAA using samples of one year $\bm{\Delta f}_d$ are shown.}
	\label{fig:bias_days_years}
\end{figure}

However, simple linear extrapolation does not work in (\ref{eq:SAA_degr}) as the degradation $\Delta C^{k}$ is a nonlinear function of $\bm{\Delta f}$. 
Therefore, we approximated $\Delta C^{k}$ as follows. 
Concatenate all samples of one day $\bm{\Delta f}_d^i \in \mathcal{D}$ into $\bm{\Delta f}_\mathcal{D} = (\bm{\Delta f}_d^1 , \bm{\Delta f}_d^2, \ldots, \bm{\Delta f}_d^{n_\mathcal{D}}) \in \mathbb{R}^{n_{\mathcal{D}} n_{t,d}}$ and simulate the BESS model (\ref{eq:BESS_model}) to receive the corresponding $\bm{SoC}_\mathcal{D} \in \mathbb{R}^{n_{\mathcal{D}} n_{t,d}}$. Using the rainflow counting algorithm (\ref{eq:rainflow}), one receives the corresponding $SoC_{av,i_c}^{cyc}, DoD_{i_c},$ $Q_{i_c} = Rainflow(\bm{SoC}_\mathcal{D} )$ of each cycle $i_c = 1,\ldots, n_{cyc}$. As this accounts for only $n_{\mathcal{D}}$ days out of the 365 in a complete year, we obtained an approximation of the degradation due to cycling by scaling the throughput with $365/n_{\mathcal{D}}$. To approximate the calendar ageing, we used the empirical mean over $\bm{SoC}_\mathcal{D}$ as $SoC_{av}^{cal}$ in~(\ref{eq:CapDegr}):



\begin{subequations}\label{eq:degr_days_appr}
\begin{align}
\E[\Delta C^{cyc,k}(\bm{x}, \bm{\Delta f})] & \approx \Delta C_\mathcal{D}^{cyc,k}(\bm{x}, \bm{\Delta f}_\mathcal{D}) = - \sum_{i_c=1}^{n_{cyc}}  \beta_{cap}(SoC_{av, i_c}^{cyc},DoD_{i_c}) \cdot \left( \sqrt{\frac{365}{n_{\mathcal{D}}} Q_{i_c} } - \sqrt{\frac{365}{n_{\mathcal{D}}} Q_{i_c-1}} \right),  \label{eq:degr_days_appr_cyc}\\
\E[\Delta C^{cal,k}(\bm{x}, \bm{\Delta f})] & \approx \Delta C_\mathcal{D}^{cal,k}(\bm{x}, \bm{\Delta f}_\mathcal{D}) = -  \alpha_{cap}(\overline{\bm{SoC}_{\mathcal{D}}},T) \left(\left(365(k+1)\right)^{0.75} -  \left(365k\right)^{0.75} \right). \label{eq:degr_days_appr_cal}
\end{align}
\end{subequations}

To illustrate the quality of this approximation, Figure \ref{fig:bias_days_years} shows a boxplot of the relative error of the SAA in (\ref{eq:opt_problem_obj}) using samples of one day $\bm{\Delta f}_d$ as explained in the paragraph above, compared to the SAA in (\ref{eq:opt_problem_obj}) using samples of one year $\bm{\Delta f}_y$, for various numbers of samples of one day $n_\mathcal{D}$. One can see that the SAA using $\bm{\Delta f}_d$ converges towards the expected value of the SAA using $\bm{\Delta f}_y$, for $n_\mathcal{D} \rightarrow \infty$. The expected value of the SAA using $\bm{\Delta f}_d$ has a negligible bias of around \SI{0.3}{\%}. The figure also shows the \SI{99}{\%} Monte Carlo confidence intervals of the expected value when using samples of one day and when using four samples of one year of frequency data, which equals $4\times365=1460$ samples of one day. 


\subsubsection{Evaluation of SAA Solution Quality}

\begin{figure}
	\centering
	\includegraphics[width=0.4\textwidth]{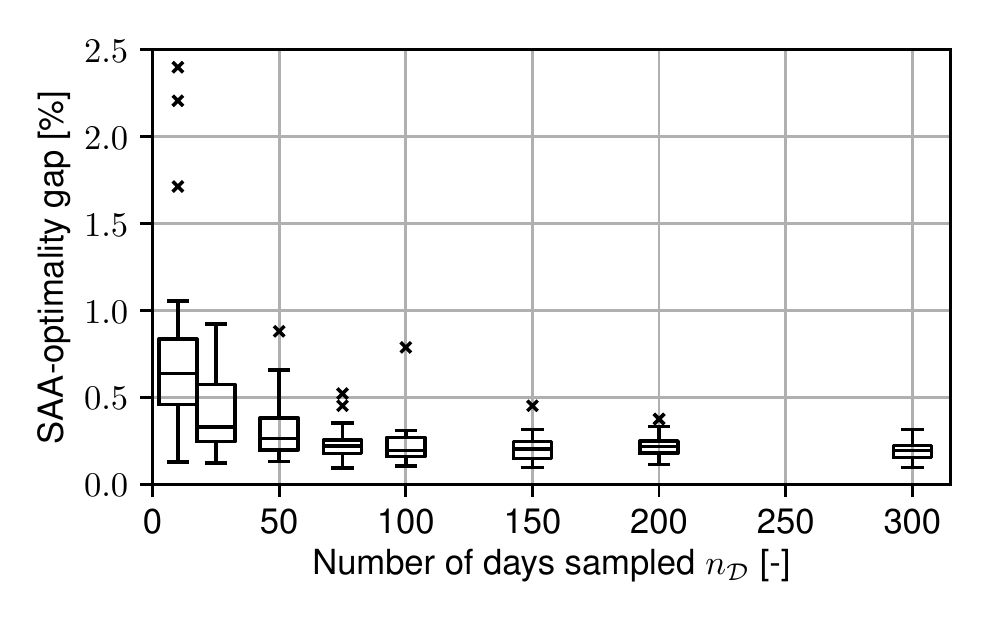}
	\caption{Boxplot of the \SI{99}{\%}-confidence optimality gap due to the SAA approximation of (\ref{eq:opt_problem}) for various sample sizes $n_\mathcal{D}$ used in the optimisation. For each sample size, the problem was solved 25 times with different sample sets $\mathcal{D}$.}
	\label{fig:montecarlo_error}
\end{figure}

To estimate the amount of samples $n_\mathcal{D}$ needed, we evaluated the SAA quality of a candidate solution $\hat{\bm{x}}$ obtained by the optimisation algorithm presented further, for various sample sizes $n_\mathcal{D}$.
To evaluate the SAA quality of $\hat{\bm{x}}$, we used the approach of Mak et al.~\cite{Mak1999}. Let $g(\bm{x}, \bm{\Delta f}_y)$ be the value of objective function (\ref{eq:opt_problem_obj}) evaluated at $\bm{x}$ with frequency sample $\bm{\Delta f}_y$ and define

\begin{equation}\label{eq:gap}
G_{n_\mathcal{Y}}^i = \frac{1}{n_\mathcal{Y}}\sum_{\bm{\Delta f}_y \in \mathcal{Y}^i} g(\hat{\bm{x}}, \bm{\Delta f}_y) - \min_{\bm{x}\in \mathcal{X}} \frac{1}{n_\mathcal{Y}} \sum_{\bm{\Delta f}_y \in \mathcal{Y}^i} g(\bm{x}, \bm{\Delta f}_y), 
\end{equation}
with $\mathcal{Y}^i, |\mathcal{Y}^i| = n_\mathcal{Y}$ being a set of iid frequency samples of one year, then $\E[G_{n_\mathcal{Y}}] \geq gap_{SAA}(\hat{\bm{x}})$ is the optimality gap due to the SAA method at $\hat{\bm{x}}$. Therefore, by sampling $n_g$ batches $\mathcal{Y}^0, \ldots, \mathcal{Y}^{n_g}$ and calculating $G_{n_\mathcal{Y}}^i, i = 1,\ldots, n_g$, we can obtain a $100(1-\beta)\%$ confidence bound on the SAA optimality gap from $1/n_g \sum_i^{n_g} G_{n_\mathcal{Y}}^i + s (G_{n_\mathcal{Y}}) t_{\beta, n_g-1} / \sqrt{n_g}$, with $t_{\beta, n_g-1} $ being the $\beta$-percentile of the Student's $t$-distribution with $n_g-1$ degrees of freedom and $ s (G_{n_\mathcal{Y}})$ being the sample standard deviation of $G_{n_\mathcal{Y}}^i$.

Figure \ref{fig:montecarlo_error} shows this \SI{99}{\%}-confidence SAA optimality gap for various sample sizes $n_\mathcal{D}$ used in the optimisation routine. The resulting solutions $\hat{\bm{x}}$ were evaluated using (\ref{eq:gap}) with $n_\mathcal{Y}=3$, i.e. sampling three years out of the four years of available data with replacement, and $n_g=20$. For each sample size $n_\mathcal{D}$, the optimisation was performed 25 times to obtain a view on the statistics of the SAA optimisation gap. As can be seen in the figure, the SAA optimality gap decreased quickly to be $<\SI{1}{\%}$ and followed a $1/n_\mathcal{D}$ trend, which is as expected when using an SAA. 

Note that the actual optimality gap at $\hat{\bm{x}}$ consists of two parts: the optimality gap due to the SAA $gap_{SAA}(\hat{\bm{x}})$ discussed here and an optimality gap due to the heuristic optimisation algorithm elaborated in Section \ref{sec:opt_algo}, whose global optimality cannot be proven.

\subsection{Chance Constraint Approximation}
The optimisation problem (\ref{eq:opt_problem}) is constrained by a chance constraint (\ref{eq:opt_problem_constr}), which limits the probability of penalties due to bad delivery of the frequency control service to be below a threshold $\epsilon^{req}$. 
Generally, chance constraints are dealt with by one of the following two methods:
The first method is to use an analytical reformulation of (\ref{eq:opt_problem_constr}), which is not possible in our case owing to the unavailability of a closed mathematical form of the BESS model. Moreover, a realistic stochastic model of $\bm{\Delta f}$ using analytical distributions is difficult to set up as it concerns a very high-dimensional multivariate stochastic variable.

A second method is to use Monte Carlo sampling to approximate the value of the probability of (\ref{eq:opt_problem_constr}). Scenario methods~\cite{Calafiore2006} provide explicit bounds on the number of samples one needs to constrain. However, these are only valid for convex optimisation problems. Statistical learning theory~\cite{Vapnik1998, VIDYASAGAR2001} is applicable to non-convex control design; however, it requires a very large number of samples~\cite{Grammatico2016} and the method uses the Vapnik--Chervonenkis (VC) dimension~\cite{VapnikChervonenkis1971}, which is very difficult to compute for general functions and can be infinite. 

As neither method is practically applicable to our model, the most we can do is to perform an a posteriori evaluation of a candidate solution $\hat{\bm{x}}$ in (\ref{eq:opt_problem_constr}) using the classical Monte Carlo approach as follows.
Consider $n_c$ iid frequency samples $\bm{\Delta f}_i$, and let 
$m = \sum_i^{n_c} \mathbbm{1}\{ p^k(\hat{\bm{x}},\bm{\Delta f}_i)>0\}$ 
be the total number of constrained violations, i.e. the number of times a frequency sample induces a penalty. A $100(1-\beta)\%$ confidence upper bound to the probability of (\ref{eq:opt_problem_constr}) is then given by \cite{Shapiro2009}:
\begin{equation}\label{eq:chance_ub}
\Pr\{ p^k(\hat{\bm{x}}, \bm{\Delta f}) > 0 \} \leq \sup_{\rho\in [0,1]} \{\rho: \mathfrak{b}(m; \rho, n_c)\geq \beta  \} \leq \epsilon^{req},
\end{equation}
with $\mathfrak{b}(m; \rho, n_c)$ being the cumulative binomial probability function with parameters $n_c$ and $\rho$, evaluated at $m$. As  $\mathfrak{b}(m; \rho, n_c)$ is continuous and monotonically decreasing in $\rho \in (0,1)$, the supremum $\sup_{\rho\in [0,1]}$ can easily be calculated by, e.g. a line search along $\rho$.

Expression (\ref{eq:chance_ub}) also defines the maximum number of samples with a penalty $m_{max}$ one can allow to ensure that (\ref{eq:opt_problem_constr}) is true with a confidence of $100(1-\beta)\%$. For instance, if one requires $\epsilon^{req} \leq \rho \leq 0.005, \beta=0.001$, and one uses $n_c = 10000$ samples, then $m\leq m_{max} = 29$.



\subsection{Optimisation Algorithm}\label{sec:opt_algo}
Even with the approximations explained above, the optimisation problem (\ref{eq:opt_problem}) is non-convex and a closed mathematical form of (\ref{eq:opt_problem}) is not readily available, resulting in an intractable problem.
However, given a parameter vector $\bm{x}$ and a frequency sample $\bm{\Delta f}^i$, one can simulate the BESS model (\ref{eq:BESS_model}) with the FCR controller (\ref{eq:FCR_controller}) and calculate the corresponding degradation using (\ref{eq:Degr}) and penalties $p^k$ quite efficiently.
This allowed us to employ a global optimisation algorithm that only requires function evaluations to find an approximate solution, without, however, providing any optimality guarantees. 

To solve (\ref{eq:opt_problem}), we propose the use of the differential evolution algorithm~\cite{Storn1997}, which belongs to the family of genetic algorithms. This has the advantage of being gradient free, which is required as the objective and constraints are non-differentiable. 
Although other gradient-free global optimisation algorithms can also be applied, we found that the differential evolution converges relatively fast 
and consistent towards a good suboptimal solution.


As constraints cannot be enforced directly in these types of global optimisation algorithms, we incorporated the constraint (\ref{eq:opt_problem_constr})  into the objective with an if-condition, returning a term proportional to the maximum penalty of a frequency dataset used for penalty checking $\mathcal{P}\subset \mathbb{R}^{n_{t,d}}, |\mathcal{P}| = n_{\mathcal{P}}$, if there is indeed a penalty detected in this frequency set.
The optimisation objective can then be written down as
\begin{equation}
g^k(\bm{x},\mathcal{D},\mathcal{P}) = 
\begin{cases}
-\E[c_{FCR}^k] r +  \frac{365}{n_\mathcal{D}} \sum_{\bm{\Delta f}_d \in \mathcal{D}}  c_{elec}^k(\bm{x}, \bm{\Delta f}_d) + \frac{(\ref{eq:degr_days_appr_cyc}) + (\ref{eq:degr_days_appr_cal})}{\SI{100}{\%}-\SI{80}{\%}}c_{cell}
& \text{if } \max_{\bm{\Delta f}\in\mathcal{P}} p^k(\bm{x}, \bm{\Delta f}) \leq 0 \\
c_{p} \max_{\bm{\Delta f}\in\mathcal{P}} p^k(\bm{x}, \bm{\Delta f})
& \text{otherwise}, 
\end{cases}
\label{eq:opt_problem_final}
\end{equation}
with $c_{p}$ being a weighting factor.

The optimisation objective thus depends on two sets of frequency samples: $\mathcal{D}, \mathcal{P} \subset \mathbb{R}^{n_{t,d}}$; the former consists of iid samples for the SAA of (\ref{eq:opt_problem_obj}), whereas the latter is a set of samples used to check the violation of the penalty constraint. 
From Figure \ref{fig:montecarlo_error}, we can observe that selecting $n_\mathcal{D} = 50$ results in an SAA optimality gap of $<1\%$. The set $\mathcal{P}$ can be thought of as a \emph{worst-case} frequency dataset containing extreme samples that the BESS should be able to provide without incurring penalties. This set will be generated during the optimisation algorithm, which is shown in Algorithm \ref{alg}.

\begin{algorithm}
	\caption{Optimisation algorithm}
	\label{alg}
	\begin{algorithmic}[1]
	\State $C^0 \gets C^{init}, R^0_0 \gets R^{init}_0, R^0_1 \gets R^{init}_1, k \gets 0$.
	\State $\mathcal{D} \gets \{\bm{\Delta f}_d^1, \ldots, \bm{\Delta f}_d^{n_\mathcal{D}} \}$ iid frequency samples.
	\While{$C^k \geq 0.8 C^{init}$ \textbf{and} $\epsilon^{k-1}\leq \epsilon^{req}$,}
		\State $i \gets 0, \bm{x}_i \gets \bm{x}^{init}, \mathcal{P} \gets \emptyset, n_{check} \gets n_{check}^{init}$.
		\While{$not(StoppingCriterion)$,}\label{alg:stop}
		\State $i \gets i+1$.
		\State 
		$\bm{x}_{i} \gets DifferentialEvolutionStep \text{ with } g^k(\bm{x},\mathcal{D}, \mathcal{P})$, according to \cite{Storn1997}. \label{alg:diffevolv}
		\If{$i = n_{check}$}\label{alg:startcheck}
			\State $m \gets \sum_i^{n_c} \mathbbm{1}\{ p^k(\bm{x}_i,\bm{\Delta f}_d^i)>0\}$, with $\bm{\Delta f}_d^1,\ldots, \bm{\Delta f}_d^{n_c} \in \mathbb{R}^{n_{t,d}}$ drawn iid.
			\If{$\sup_{\rho\in [0,1]} \{\rho: \mathfrak{b}(m; \rho, n_c)\geq \beta \} > \epsilon^{req}$}
				\State Sort $\bm{\Delta f}_d^i$, so that $ p^k(x_i,\bm{\Delta f}_d^{(1)})\leq p^k(\bm{x}_i,\bm{\Delta f}_d^{(2)})\leq \ldots \leq p^k(x_i,\bm{\Delta f}_d^{(n_c)}) $. \label{alg:sort} 
				\State $\mathcal{P} \gets \mathcal{P} \cup \bm{\Delta f}_d^{(j^*)} $, with $j^*=n_c-m_{max}$. \label{alg:add} 
				\State $n_{check} \gets n_{check}^{init}$. \label{alg:ncheck}
			\Else
				\State $n_{check} \gets n_{check} + n_{check}/2$. 
			\EndIf\label{alg:stopcheck}
		\EndIf
		\EndWhile
	\State $\epsilon^k \gets  \sup_{\rho\in [0,1]} \{\rho: \mathfrak{b}(m'; \rho, n'_c)\geq \beta  \}$, with $n'_c > n_c$.\label{alg:prob}
	\State Get $\bm{SoC}_y$ by simulating the BESS model (\ref{eq:BESS_model}),(\ref{eq:FCR_controller}) with $\bm{x}_i,$ $\forall 
	\bm{\Delta f}_y \in \mathcal{Y}$. \label{alg:socy}
	\State $C^{k+1} \gets \overline{C^{k+1}}, R^{k+1}_0\gets \overline{R^{k+1}_0}, R^{k+1}_1\gets \overline{R^{k+1}_1}$, using (\ref{eq:Degr})--(\ref{eq:cycle_degr}) with $\bm{SoC}_y, \forall y =1\ldots n_{\mathcal{Y}}$. \label{alg:capupdate}
	\State $k\gets k+1$.
	\EndWhile
	\State $k_{max} \gets k-1$. \label{alg:end}
	\end{algorithmic}
\end{algorithm}

Steps \ref{alg:startcheck} to \ref{alg:stopcheck} in Algorithm \ref{alg} show the procedure used to create set $\mathcal{P}$: for a given point $\bm{x}_i$, draw $n_c$ iid samples and calculate the upper bound on constraint violation using (\ref{eq:chance_ub}). 
If this upper bound is higher than the required $\epsilon^{req}$, one cannot ensure that (\ref{eq:opt_problem_constr}) is satisfied with confidence $\beta$ for the current most optimal point $\bm{x}_i$. 
We know from (\ref{eq:chance_ub}) that there can be maximum $m_{max}$ samples with a penalty $p^k>0$ given $n_c$. Thus, in steps \ref{alg:sort} and \ref{alg:add}, we add the $\Delta f_d^{(j^{*})}$ sample that gives the $j^{*}th$-largest penalty, with $j^{*}=n_c-m_{max}$, to $\mathcal{P}$ as this is the sample with the largest penalty of all samples that are actually not allowed to have a penalty at all.
If the upper bound (\ref{eq:chance_ub}) is smaller than or equal to $\epsilon^{req}$, the optimisation continues with the same $\mathcal{P}$ as before and one is guaranteed that (\ref{eq:opt_problem_constr}) is satisfied with confidence $100(1-\beta)\%$ at $\bm{x}_i$.

As evaluating the penalty $p^k$ on $n_c$ samples is computationally expensive owing to the large number of samples required for small $\epsilon^{req}$, we only check this after $n_{check}$ iterations. If no penalties are found, $n_{check}$ is updated according to an exponential update rule in step \ref{alg:stopcheck}.

In step \ref{alg:diffevolv}, the differential evolution performs one optimisation step, in which it updates each member of its population and returns the population member with the lowest objective value $g^k(x,\mathcal{D}, \mathcal{P})$. In this study, we used a population size of 60 members, chose the best member to be mutated and used a binomial crossover scheme. The differential evolution stops if the standard deviation of the objective values of the population is smaller than $ 5\cdot10^{-4}$ times the mean of the objective values of the population. 

When converged to an optimal value $\hat{\bm{x}}$, we check the actual probability on a penalty in step \ref{alg:prob} on a broader set $n'_c>n_c$ of iid samples and calculate the empirical mean of the capacity degradation $\overline{C^{k+1}}$ and resistance growth $\overline{R^{k+1}_0}, \overline{R^{k+1}_1}$ over all available years in the dataset $\mathcal{Y}$, which serves as the capacity and resistance for year $k+1$. The algorithm stops when the battery capacity reaches \SI{80}{\%} of its initial capacity or when it is unable to provide the service, with the probability on a penalty being smaller than required (i.e. $\epsilon^k > \epsilon^{req}$).

\subsection{Total Revenues and Costs}
Algorithm~\ref{alg}, gives a solution to optimisation problem~(\ref{eq:opt_problem}) for each year $k$ the BESS is able to deliver FCR services.
To then calculate the total expected revenues and costs of the BESS over its lifetime, we use the resulting optimised control variables $\hat{\bm{x}}^k$ and expected capacity degradation $ \overline{C^{k+1}}$ of the optimisation routine defined by Algorithm~\ref{alg}.
The expected electricity costs of year $k$ can then be estimated by taking the empirical mean over all frequency samples of one year $\hat{c}_{elec}^k = \overline{c_{elec}^k(\hat{\bm{x}}^k,\bm{\Delta f}_y)}$. The FCR revenues of year $k$ are simply the product of $r$ and $\E[c_{FCR}^k]$, except for the last year $k_{max}-1$.
To evaluate the proportion of year $k_{max}-1$ the battery is still able to provide the service, we perform a linear interpolation.
The total discounted net revenues of the BESS can then be calculated as
\begin{equation}\label{eq:rev}
rev = \sum_{k=1}^{k_{max}} \frac{\E[c_{FCR}^k] r - \hat{c}_{elec}^k }{(1+\gamma)^k} \cdot \max \left(\min \left(\frac{0.8-C^{k-1}}{C^k-C^{k-1}}, \frac{\epsilon^{req}-\epsilon^{k-1}}{\epsilon^k-\epsilon^{k-1}}, 1\right),0\right),
\end{equation}
with $k_{max}$ as determined by step \ref{alg:end} of the optimisation algorithm and $\gamma$ being an appropriate discount rate. As long as the battery is not degraded in year $k$ ($C^k > 0.8$), and is able to provide the FCR services with a probability on a penalty smaller than required ($\epsilon^k > \epsilon^{req}$), the second term in the equation will equal to one and the FCR revenues of the entire year are taken into account. 
If this is not the case, the minimum in the second term is taken between the linear interpolation of the degradation and the linear interpolation of the $\epsilon^{req}$ metric in case both occur during the same year $k$. 

\section{Case Study: BESS in German FCR}\label{sec:results}
In this section, we discuss the application of the proposed optimisation algorithm to the German FCR market, which is currently the largest FCR market in Europe and has a considerable amount of BESS capacity participating. 
In Germany, FCR is auctioned through the common platform \emph{Regelleistung}~\cite{Regelleistung} shared by the four German TSOs (TenneT, Amprion, 50Hertz and TransnetBW). Starting from 2012, TSOs of neighbouring countries have been coupling their primary frequency control markets to the Regelleistung platform, which currently manages the joint tendering of FCR volume for the German, Swiss, Austrian, Belgian, French and Dutch TSOs. 

On Regelleistung, each week, a week-ahead auction is organised, where the bids are placed in merit order and the market clears on the price of the bid, where the cumulative sum of the bid volumes equals the tendering volume, subject to certain cross-border constraints. 
The market is a pay-as-bid market, which means that each selected participant is awarded its bid price rather than the marginal clearing price. However, the bidcurves are generally relatively flat, meaning that participants are fairly good in forecasting the marginal price.

In the case study, we applied all relevant regulations as they are currently imposed. We used real, measured frequency data from the CE region together with the detailed BESS model elaborated in Section~\ref{sec:model_meth} to obtain results that are practically relevant in real applications.



\subsection{Data and Regulatory Requirements}\label{sec:regulatory_req}
Regulation in Germany on battery storage in frequency control reserve is one of the most detailed of the entire ENTSO-E region. Specific requirements for batteries are given in \cite{EckpunkteFreiheitsgrade, Speicherkapazitat} and will be discussed shortly in the following paragraph.
\subsubsection{BESS in German FCR}\label{sec:GermanFCR_BESS}
When providing FCR in Germany, one is restricted to the degrees of freedom described in \cite{EckpunkteFreiheitsgrade}. The maximum frequency deviation at which all FCR power should be active is $\Delta f_{max} = \SI{200}{mHz}$ and should be reached after \SI{30}{s}. There is a \SI{10}{mHz} deadband in which no delivery is required. 

Overdelivery is allowed, but only up to \SI{20}{\%}, limiting the parameter $o_d \in\left[0.0,0.2\right]$. Recharging the battery is allowed by reserving BESS power capacity that cannot be sold as FCR power. One also has to compensate the recharging power with other assets or by buying/selling the power on the intraday markets. 
The latter option presents the additional constraints that the recharging power should be constant for 15 min, corresponding to the intraday trading blocks, and be decided upon with a lead time of at least 5 min, as the German intraday market closes 5 min before delivery. 
Moreover, as the granularity on the intraday market is \SI{100}{kW}, the recharge power has to be constrained to multiples of \SI{100}{kW}~\cite{IntradayGE}. 

When providing FCR with any energy-constrained asset in Germany, one has to comply with the \emph{30-min criterion}~\cite{Speicherkapazitat}. This criterion states that, except in emergency states, the battery should always reserve enough energy to be able to provide 30 min of FCR power in both the positive and the negative direction. An emergency state is reached if $|\Delta f|> \SI{200}{mHz}$, $|\Delta f| > \SI{100}{mHz}$ for longer than 5 min or $|\Delta f| > \SI{50}{mHz}$ for longer than 15 min. This also implies that, in case $\SI{10}{mHz} < |\Delta f| \leq \SI{50}{mHz}$, the battery should be able to deliver for an infinite amount of time. This is only possible if the recharge power is able to compensate for the required delivery, resulting in the additional requirement that $P^{rech}_{max} \geq 0.25 r$, the power sold as FCR capacity, or $r\leq 0.80P_{max}^{BESS}$.

When prequalifying the battery to participate in the Germany FCR market, one has to perform a \emph{Doppelh\"{o}ckertest}~\cite{Speicherkapazitat}, which is used to determine the available energy capacity of the battery at the FCR power $r$ one wants to prequalify. The test starts at full SoC and consists of two times a discharge period of 15 min at the FCR power $r$ followed by a rest period of 15 min. After this, the BESS has to discharge further at FCR power $r$ until the battery is empty. The total discharged energy is used to monitor the 30-min criterion defined above.

\subsubsection{Electricity Costs}
BESSs in Germany are exempt from many grid costs and other levies that typically apply to the regular consumer. An overview of all elements making up cost of the electricity is given in Table \ref{tbl:gridcost_GE}. The battery is exempted from network charges and electricity tax, and pays the EEG (Renewable Energy Resources Act) and KWK (combined heat and power) levies only on the losses incurred in the BESS. We assumed that the battery buys and sells its recharge power on the intraday market, whereas the remaining energy is settled on the imbalance price. All other costs and levies only have to be paid on the energy consumed from the grid. We assumed electricity costs rise with inflation.

\begin{table}[h]
\centering

\begin{tabular}{ l r >{\raggedright}p{2.7cm} p{4cm}}
	\toprule
	Cost Element & Amount & Applicable & Regulation and Law\\ \midrule
	Recharge power & Intraday market & Applicable & \cite{EckpunkteFreiheitsgrade} \\
	FCR power & Imbalance market & Applicable & -- \\
	Network charges & --  & Exempted & EnWG \S118 Abs. 6~\cite{EnWG} \\
	Electricity tax & \SI{2.05}{c\EUR/kWh} & Exempted & StromStG \S5 Abs. 4 \cite{StromStG} \\
	EEG levy & \SI{6.88}{c\EUR/kWh} & Exempted except for losses & EEG \S61k\cite{EEG} \\

	KWK levy & \SI{0.4438}{c\EUR/kWh} & Exempted except for losses & KWKG \S27b~\cite{KWKG} \\
	StromNEV \S19-Levy & \SI{0.370}{c\EUR/kWh} & Applicable & StromNEV \S19\cite{StromNEV}  \\
	Concession Fee & 0.11-\SI{2.39}{c\EUR/kWh} & Applicable & KAV \S1-2 \cite{KAV} \\
	Offshore liability levy & \SI{0.037}{c\EUR/kWh} & Applicable & EnWG \S17(f)~\cite{EnWG}\\
	Interruptible load levy & \SI{0.011}{c\EUR/kWh} & Applicable & AblaV \S1 \cite{ablav} \\
	\bottomrule
\end{tabular}
\caption{Elements making up the cost of electricity in Germany for a grid-connected standalone battery.}
\label{tbl:gridcost_GE}
\end{table}

\subsubsection{FCR Price}

\begin{figure}[t]
	\centering
	\includegraphics[width=0.45\textwidth]{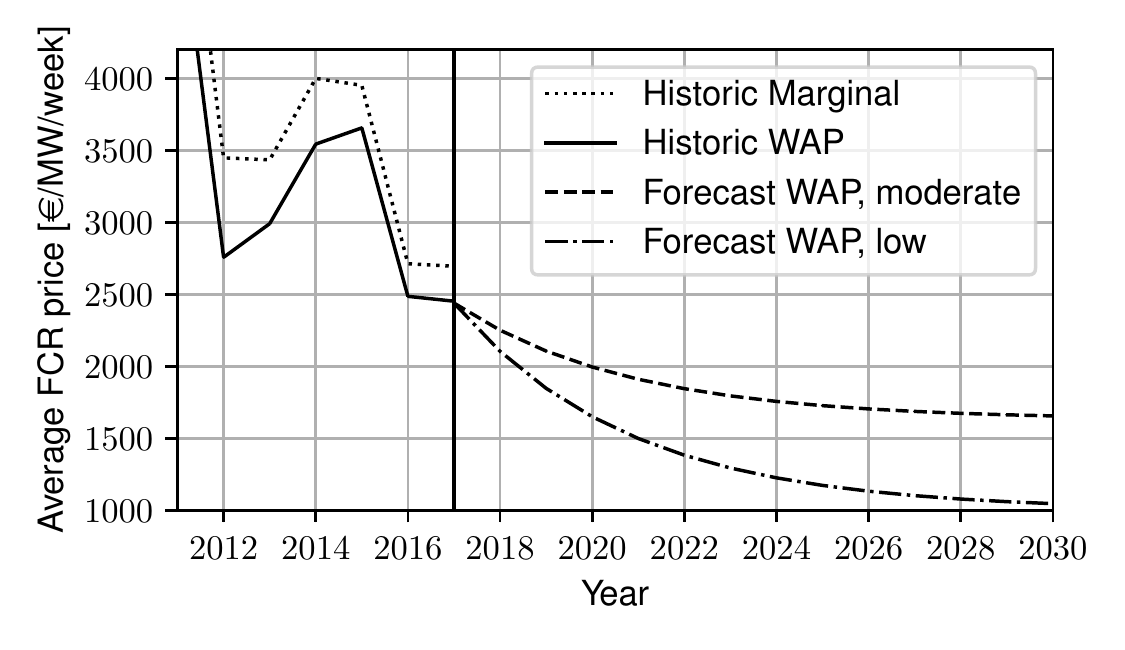}
	\caption{Historical marginal and weighted average price (WAP) in the German FCR market averaged per year~\cite{Regelleistung}. As from 2017, a moderate and a low WAP forecast scenario are shown (nominal value).}
	\label{fig:PRLprices}
\end{figure}

Figure~\ref{fig:PRLprices} shows the yearly averaged historical FCR prices published on the Regelleistung platform~\cite{Regelleistung}. The figure shows both the yearly averaged marginal price and the weighted average accepted bid price (WAP). 
As the FCR volume auctioned on the Regelleistung platform is much larger than that of a single regular-sized BESS, we assume the BESS operates as a price-taker in the FCR market.
We assume a bidding strategy that is able to capture the WAP rather than the marginal price. 
In the literature, the FCR price of the German market is forecasted to decrease in the coming years, mainly due to an increase of BESSs in the market that can provide the service at lower costs~\cite{Fleer2018}. Various predictions can be found, varying from a value of \SI{1630}{\EUR/MW/week}~\cite{Kopiske2017} to below \SI{1000}{\EUR/MW/week}~\cite{Fleer2017,Fleer2018} in 2035. We adopted two exponentially decreasing scenarios, a moderate and a low scenario, both shown in Figure~\ref{fig:PRLprices}, which correspond to the forecasts from~\cite{Fleer2017,Fleer2018, Kopiske2017}. 


\subsection{Optimisation Setup}\label{sec:opt_setop}
In optimizing and evaluating the model, we used four years (2014-2017) of frequency, intraday and imbalance market data ($n_\mathcal{Y}=4$). 
We split these data into samples of one day, starting from each quarter hour in the dataset, which resulted in 140\,256 samples of one day.
We discretised the model with a time step $\Delta t = \SI{10}{s}$. The parameters of the battery cell are as given in Table \ref{tbl:Parameters}. 
To determine the number of cells of a BESS with a certain rated energy capacity $E^{BESS}_{rated}$, we used the rated energy capacity of one cell $E_{rated}$, given in Table \ref{tbl:Parameters}, which is equal to the nominal capacity $C$ times the nominal voltage $V_{nom}$. 
Although the actual energy capacity varies with the rated power of the BESS, as explained in Section \ref{sec:E_P_cap_BESS}, this approach accounts for an easy one-to-one relation of energy capacity to the number of cells, rather than relying on more complex relationships.
For example, a BESS rated at $E^{BESS}_{rated} = \SI{1}{MWh}$ would then contain $\SI{1 000 000}{Wh}/\SI{7.38}{Wh} = \SI{135 501}{cells}$.

This also means that, when executing the  \emph{Doppelh\"{o}ckertest}~\cite{Speicherkapazitat}, which is used to determine the energy boundaries for the 30-min criterion, the actual energy discharged will differ from the rated energy capacity $E_{rated}^{BESS}$. Therefore, in our model, we simulated the Doppelh\"{o}ckertest to determine the maximum and minimum state of charge boundaries $SoC^{max,k}_{30min}$ and $SoC^{min,k}_{30min}$, and used these to determine the penalty metric $p^k$:
\begin{equation}\label{eq:penalty}
p^k = \frac{1}{n_t} \sum_t^{n_t}{ \left( \mathbbm{1}\{ SoC_t > SoC^{max,k}_{30min}\}  + \mathbbm{1}\{ SoC_t < SoC^{min,k}_{30min}\}\right) S^{emerg}_t}, 
\end{equation}
where $S^{emerg}_t$ is $0$ if $t$ is in an emergency state and $1$ otherwise,  and $n_t$ denotes the number of time steps in the sample. The dependency of  $SoC^{max,k}_{30min}$ and $SoC^{min,k}_{30min}$ on the year $k$ is due to the degradation in capacity $C^k$ and growth of resistances $R_0^k, R_1^k$, which also changes the actual energy capacity one can discharge during the Doppelh\"{o}ckertest.
Furthermore, we require $\epsilon^{req} = 0.005$ and $\beta = 0.001$, giving a very small probability on penalties. To achieve this, we set $n_c = 10000$ and $n_c' = 50000$. Finally, we have chosen $n_\mathcal{D}=50$, which gives a good compromise between SAA error and number of samples (see Figure \ref{fig:montecarlo_error}).

With this configuration, Algorithm \ref{alg} needs, on average, 202 iterations per year $k$, which takes around 24 min when run on a \SI{2.83}{GHz} Intel Core 2 Quad Processor (Q9550) with \SI{10}{GB} of RAM. 
To speed up the execution time, one can run the algorithm in parallel for different battery configurations.

\subsection{Results and Discussion}\label{sec:res_and_disc}
\begin{figure}
	\centering
	\hfill
	\begin{subfigure}[t]{0.49\textwidth}
		\centering
		\includegraphics[width=\textwidth]{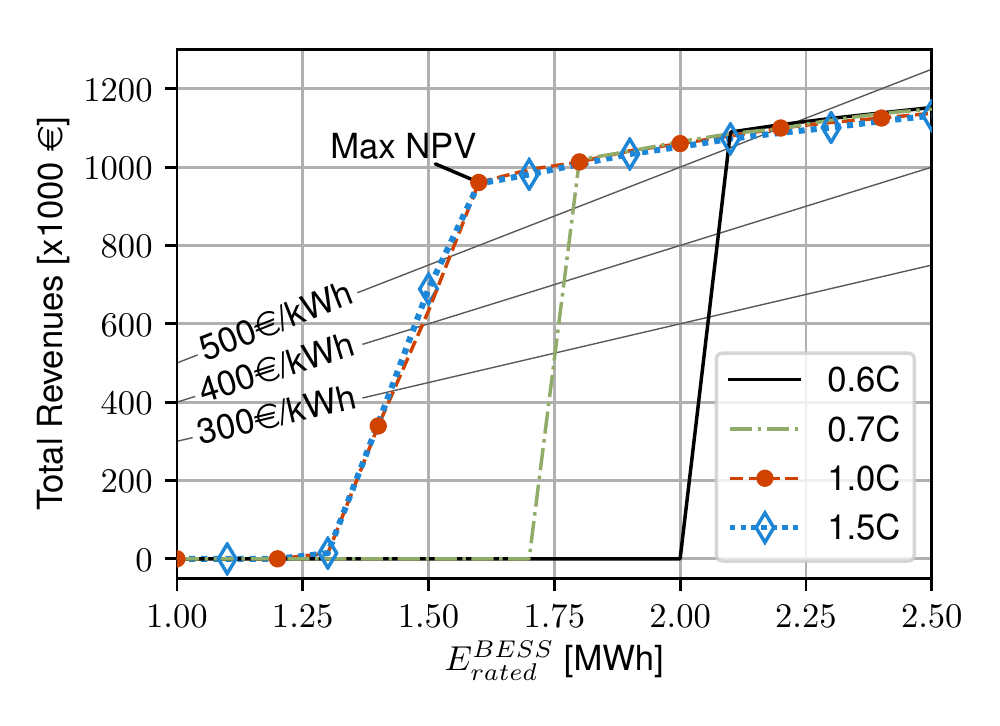}
		\caption{Moderate German FCR price scenario.}
		\label{fig:res_GE_moderate}
	\end{subfigure}
	\hfill
	\begin{subfigure}[t]{0.49\textwidth}
		\centering
		\includegraphics[width=\textwidth]{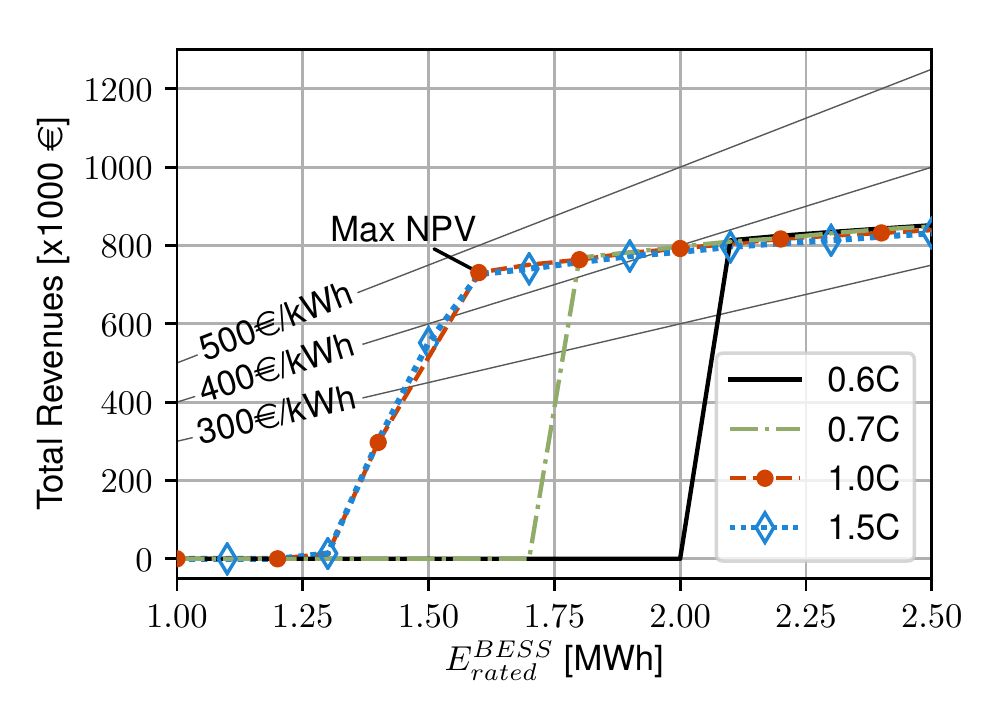}
		\caption{Low German FCR price scenario.}
		\label{fig:res_GE_low}
	\end{subfigure}
\hfill
	\caption{Total revenues over the lifetime of a BESS providing FCR services to the German market minus the electricity costs (\ref{eq:rev}), optimised according to Algorithm \ref{alg}, in the function of the rated energy capacity $E_{rated}^{BESS}$  and the C-rate. Indicative lines of battery costs in euro per kilowatt hour (\EUR/kWh) are also plotted. Revenues were adjusted for the expected inflation, assumed at $\gamma = \SI{1.7}{\%}$~\cite{InflationGE}. The point with the highest NPV, at an energy capacity of \SI{1.6}{MWh} and C-rate $\geq$ \SI{1.0}{C}, is denoted on both figures by ``Max NPV''.}
	\label{fig:results_GE}
\end{figure}

\begin{table}[h]
	\centering
	\footnotesize
	\setlength\tabcolsep{1.0pt}
	\begin{tabular}{l|*{16}{E}}
		\toprule
		& \multicolumn{16}{c}{Rated Energy Capacity $E_{rated}^{BESS}$ [MWh]}\\ 
		C-rate & 1.0   & 1.1  & 1.2  & 1.3  & 1.4  & 1.5  & 1.6  & 1.7  & 1.8  & 1.9  & 2.0    & 2.1 & 2.2 & 2.3 & 2.4 & 2.5 \\ \midrule
		\multicolumn{17}{l}{\hspace{7pt}\SI{500}{\EUR/kWh} } \\
		\SI{0.6}{C} & -500 & -550 & -600 & -650 & -700 & -750 & -800 & -850 & -900 & -950 & -1000 & 39 & 8   & -26 & -61 & -98  \\ 
		\SI{0.7}{C} & -500 & -550 & -600 & -650 & -700 & -750 & -800 & -850 & 119  & 91   & 65    & 36 & -1  & -30 & -63 & -102 \\
		\SI{1.0}{C} & -500 & -550 & -600 & -636 & -361 & -117 & 161  & 144  & 113  & 92   & 60    & 30 & 0   & -36 & -75 & -113 \\
		\SI{1.5}{C} & -500 & -550 & -600 & -636 & -354 & -61  & 158  & 131  & 108  & 84   & 52    & 22 & -12 & -49 & -82 & -118 \\
		\midrule
		\multicolumn{17}{l}{\hspace{7pt}\SI{400}{\EUR/kWh} } \\
		\SI{0.6}{C} & -400 & -440 & -480 & -520 & -560 & -600 & -640 & -680 & -720 & -760 & -800 & 249 & 228 & 204 & 179 & 152 \\
		\SI{0.7}{C} & -400 &-440 & -480 & -520 & -560 & -600 & -640 & -680 & 299  & 281  & 265  & 246 & 219 & 200 & 177 & 148 \\
		\SI{1.0}{C} & -400 & -440 & -480 & -506 & -221 & 33   & 321  & 314  & 293  & 282  & 260  & 240 & 220 & 194 & 165 & 137 \\
		\SI{1.5}{C} & -400 & -440 & -480 & -506 & -214 & 89  &  318  & 301  & 288  & 274  & 252  & 232 & 208 & 181 & 158 & 132\\ \midrule
		\multicolumn{17}{l}{\hspace{7pt}\SI{300}{\EUR/kWh} } \\
		\SI{0.6}{C}  & -300 & -330 & -360 & -390 & -420 & -450 & -480 & -510 & -540 & -570 & -600 & 459 & 448 & 434 & 419 & 402 \\
		\SI{0.7}{C}  & -300 & -330 & -360 & -390 & -420 & -450 & -480 & -510 & 479  & 471  & 465  & 456 & 439 & 430 & 417 & 398 \\
		\SI{1.0}{C}   & -300 & -330 & -360 & -376 & -81  & 183  & 481  & 484  & 473  & 472  & 460  & 450 & 440 & 424 & 405 & 387 \\
		\SI{1.5}{C}  & -300 & -330 & -360 & -376 & -74  & 239  & 478  & 471  & 468  & 464  & 452  & 442 & 428 & 411 & 398 & 382 \\
		\bottomrule
	\end{tabular}
	\caption{Net present value in $\times$\SI{1000}{\EUR} at a discount rate $\gamma = \SI{1.7}{\%}$ equal to the expected inflation \cite{InflationGE} over the lifetime of a BESS providing FCR services to the German market, using the moderate FCR price scenario of Figure \ref{fig:PRLprices}, for a BESS cost $cost_{BESS}$ of \SI{500}{\EUR/kWh}, \SI{400}{\EUR/kWh} and \SI{300}{\EUR/kWh}.}
	\label{tbl:NPV}
\end{table}

The goal is to determine the optimal BESS power and energy capacity to provide $r=\SI{1}{MW}$ of FCR capacity, the minimum bid size on Regelleistung, during its lifetime. For another bid size that is a multiple of \SI{1}{MW}, the results can be scaled proportionally.

To determine the optimal BESS sizing, we ran the optimisation algorithm repeatedly for a BESS with a rated energy capacity varying between \SI{1.0}{MWh} and \SI{2.5}{MWh}, and with a varying C-rate. The C-rate is defined as the rated power divided by the rated energy capacity: $\text{C-rate}=P_{max}^{BESS}/E_{rated}^{BESS}$. The results are shown in Figure~\ref{fig:results_GE} for C-rates of \SI{0.6}{C}, \SI{0.7}{C}, \SI{1.0}{C} and \SI{1.5}{C}.

Figure \ref{fig:res_GE_moderate} shows the total revenues minus the electricity costs in the FCR market according to (\ref{eq:rev}), with $\gamma = \SI{1.7}{\%}$ to adjust for inflation~\cite{InflationGE}, over the operational lifetime of the BESS, using the FCR prices of the moderate scenario of Figure \ref{fig:PRLprices}, whereas Figure \ref{fig:res_GE_low} shows the total inflation-adjusted net revenues using the low scenario. 
The figure also shows a couple of lines indicating the potential cost of a BESS, so that the net present value (NPV), which equals the discounted net revenues (\ref{eq:rev}) minus the investment costs: $NPV = rev - cost_{BESS}$, can be read on the $y$-axis by taking the difference between the revenue and the cost lines. 
Table~\ref{tbl:NPV} shows the NPV with the moderate FCR price scenario for the various BESS energy capacities, C-rates and battery costs $cost_{BESS}$.
As can be seen in the table and in Figure \ref{fig:results_GE}, a \SI{1.6}{MWh} battery with a power rating higher than \SI{1.6}{MW} results in the highest NPV and would thus be the optimal BESS sizing for the German FCR market (at a cost of only \SI{300}{\EUR/kWh}, the \SI{1.7}{MWh}/\SI{1.0}{C} battery has a slightly higher NPV and would be the better choice).

Figure \ref{fig:results_GE} shows that a BESS with a lower C-rate (\SI{0.7}{C} and \SI{0.6}{C}) can only obtain revenues by participating in the German FCR market with a relatively high rated energy capacity. This is due to the requirement 
that $P_{max}^{BESS} \geq 1.25 r \geq \SI{1.25}{MW}$. 
For a \SI{0.6}{C} BESS, this means a minimum energy capacity of \SI{2.09}{MWh} is needed.

For a BESS with a higher C-rate (\SI{1.0}{C} and \SI{1.5}{C}), one needs at least \SI{1.3}{MWh} to be able to participate with a \SI{1}{MW} FCR capacity in the German market. This is due to the 30-min criterion, which obliges to reserve at least \SI{1}{MWh} of energy capacity for emergency states. 
A battery with a rated energy capacity of \SI{1.6}{MWh} 
has much larger revenues and NPV than those of a battery rated at \SI{1.5}{MWh}, whose revenues are already much higher than those of a battery rated at \SI{1.4}{MWh} and \SI{1.3}{MWh}.
This occurs because, in this part of the graph (between \SI{1.3}{MWh} and \SI{1.6}{MWh}, C-rate $\geq$ \SI{1.0}{C}), the revenues of a BESS are limited by the years the BESS can provide the FCR service without violating the penalty constraint (\ref{eq:chance_ub}). 
In this range, a BESS with a larger energy capacity will be able to satisfy the penalty constraint for a longer period of time and therefore obtain more revenues.  

\begin{figure}
	\centering
	\includegraphics[width=0.45\textwidth]{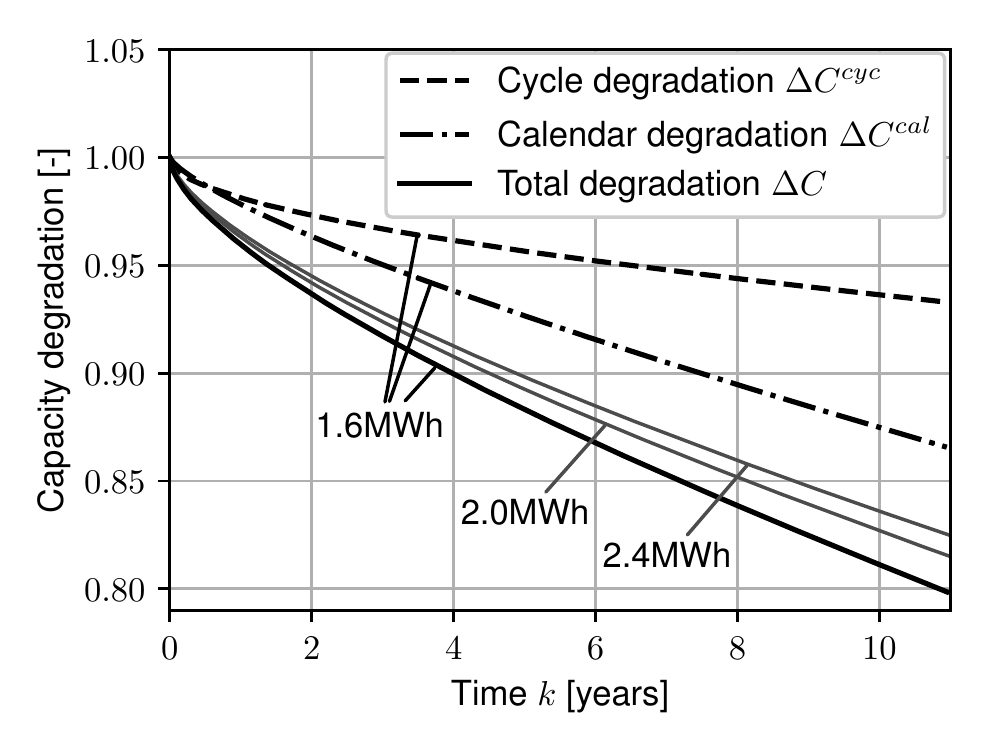}
	\caption{Total capacity degradation of a \SI{1.0}{C} BESS with a rated energy capacity of \SI{1.6}{MWh}, \SI{2.0}{MWh} and \SI{2.4}{MWh} delivering $r=\SI{1}{MW}$ of FCR power in Germany. The calendar degradation $\Delta C^{cal}$ and the degradation due to cycling $\Delta C^{cyc}$ of the \SI{1.6}{MWh}/\SI{1.6}{MW} BESS are also shown.}
	\label{fig:degradation}
\end{figure}

A battery with an energy capacity larger than \SI{1.6}{MWh} will also see larger revenues; however, the slope of the increase in revenues with larger energy capacity is smaller than the slope between \SI{1.3}{MWh} and \SI{1.6}{MWh}. 
In this part of the graph, the operational lifetime of the BESS in the FCR market is not limited by the penalty constraint, but rather by the end-of-life criterion ($C=\SI{80}{\%}$). 
The BESS will be able to provide FCR while respecting (\ref{eq:chance_ub}) its entire lifetime until it is degraded completely. 
The degradation curve of three \SI{1.0}{C} batteries is shown in Figure~\ref{fig:degradation}. 
As can be seen in the figure, a larger energy capacity reduces degradation, as the DoD of the cycles will be smaller and, therefore, extends the lifetime of the BESS.
However, the additional revenues do not outweigh the costs of installing additional energy capacity, resulting in a lower NPV with additional energy capacity, as can be seen in Table \ref{tbl:NPV}. 
It is also interesting to note that, in this part of the graph, the power capacity has almost no impact on the total revenues. This occurs because the BESS has to provide \SI{1}{MW} of FCR capacity and a BESS with a power capacity of \SI{1.6}{MW} has enough recharging power to provide the FCR service over its entire lifetime. A larger power capacity only results in a larger available recharging power, which is already sufficient to provide the FCR service, with no impact on the total revenues.

The point at \SI{1.6}{MWh} (denoted in Figure \ref{fig:results_GE} with ``Max NPV'') is where the two parts with the different slopes meet, which explains why this point has the highest NPV:
a BESS with \SI{1.6}{MWh} and a C-rate $\geq$ \SI{1.0}{C} has the smallest amount of energy capacity that is able to provide the FCR while respecting the penalty constraint (\ref{eq:chance_ub}) until it is degraded to its end-of-life criterion. The payback period of a BESS with this configuration is shown in Table~\ref{tbl:payback}, for the different costs $costs_{BESS}$ and for the moderate and low Germany FCR price scenarios.

\begin{table}[h]
	\centering
	\footnotesize
	\begin{tabular}{lrr}
		\toprule
		& \multicolumn{2}{c}{FCR price scenario}\\ 
		$cost_{BESS}$  & Moderate & Low \\ \midrule
		\SI{500}{\EUR/kWh} & 7.1 years & --  \\ 
		\SI{400}{\EUR/kWh} & 5.3 years & 7.3 years  \\ 
		\SI{300}{\EUR/kWh} & 3.6 years & 4.7 years  \\
		\bottomrule
	\end{tabular}
	\caption{Payback period of a \SI{1.6}{MWh}/\SI{1.6}{MW} BESS.}
	\label{tbl:payback}
\end{table}

The electricity costs of a BESS are very low (in the moderate scenario, between \SI{0.94}{\%} and \SI{2.97}{\%} of total FCR revenues, or between \SI{135}{\EUR} and \SI{2575}{\EUR} per year), as BESSs are exempted from many of the electricity cost elements in Germany (see Table \ref{tbl:gridcost_GE}). 
Therefore, the total net revenues are governed by the revenues of selling the FCR capacity on the market. 
As degradation is the main limitation to the amount of time a BESS can participate in the FCR market, it is interesting to take a closer look at the calendar and cycle degradation of a \SI{1.6}{MWh}/\SI{1.6}{MW} BESS, both shown in Figure \ref{fig:degradation}. 
The BESS reaches its end-of-life criterion ($C= \SI{80}{\%}$) after 10.8 years of service. As can be seen in the figure, most of the degradation is actually due to calendar degradation. 
Although there are plenty of cycles, these often have a low DoD and, thus, have a limited effect on the total degradation.

\section{Conclusions}\label{sec:conclusion}
In this paper, we have proposed a holistic, data-driven optimisation framework for the investment analysis, sizing and control design of a battery energy storage system participating in frequency control markets. 
To control the state of charge of a battery storage system performing frequency control, we used a parametrised recharge controller compliant with regulatory requirements, which we optimised to minimise degradation over the lifetime of the battery storage system using real frequency data. 

As the required activation profile when providing frequency control is stochastic, we formulated a probabilistic optimisation problem that allows the probability of being unavailable to be constrained to a small value with high confidence.
We solved the problem by adopting a global evolutionary optimisation algorithm that only requires function evaluations, which allows the use of a battery energy storage model 
of which a closed mathematical form is not directly available, but can only be simulated.
This approach allowed us to use a battery energy storage model that is more detailed than usually seen in the literature, featuring a dynamic RC battery cell model, a semi-empirical degradation model, an inverter model and an HVAC model.

Finally, we performed a techno-economic analysis of a battery in the German primary frequency control (frequency containment reserve) market, using the proposed framework. 
We considered all relevant regulations and used real frequency data, so that the results are applicable to a real case.
We found that a battery storage system rated at \SI{1.6}{MW}/\SI{1.6}{MWh} provides the highest net present value and can deliver \SI{1.0}{MW} of frequency control capacity for 10.8 years, after which it is degraded to \SI{80}{\%}, which is the end-of-life criterion. Most of the observed degradation was due to calendar degradation, as the cycles performed in frequency control had a limited depth of discharge. 

It is worth mentioning that the developed optimisation framework can easily be applied to other countries by incorporating the regulations specific to that country. Interesting future work would be to apply the developed framework to various countries and compare the impact of different regulations on the investment case of the battery system.

Although all models used in the framework are each separately validated by experiments, validating the combination of all models on a battery system providing frequency control would also be relevant future work.
Other future work may consist of extending the dynamic program (\ref{eq:opt_problem}) to incorporate a broader class of policies, different approximations of the value function or more decisions such as buying additional battery cells for the battery system or using the battery for other services could increase the total value of the battery system.
Finally, it would also be interesting to investigate the combination of frequency control with other revenue streams, such as day-ahead or imbalance market arbitraging.

\section*{Acknowledgements}
This work is supported by the Flanders Innovation \& Entrepreneurship Agency (VLAIO) (grant no. BM~160214) and by KU Leuven Research Fund (grant no. C24/16/018).

\appendix
\section{Rainflow Counting Algorithm}\label{sec:app_rain}
The rainflow counting algorithm used in (\ref{eq:rainflow}) starts from all the local extrema $\nu_i$ of the vector $\bm{SoC}$ and is shown below:

\setcounter{algorithm}{0} 
\renewcommand{\thealgorithm}{A.\arabic{algorithm}}
\begin{algorithm}
	\caption{Rainflow Counting Algorithm~\cite{cyclingStandard}}
	\label{alg: rainflow}
	\begin{algorithmic}[1]
		\State Let $\bm{\nu} = (\nu_1, \nu_2, \ldots, \nu_{n_\nu})$ be a vector containing the local extrema of $\bm{SoC} \in  \mathbb{R}^{n_t}$.
		\State $s \gets 1, i \gets 3, i_c \gets 0, Q_0 \gets  0$.
		\While{$i \leq {n_\nu}$}
		\While{$i-s < 2$}
		\State{$ i \gets i + 1$}.
		\EndWhile
		\State $\Delta_1 \gets \left|\nu_{i-2}-\nu_{i-1}\right|, \Delta_2 \gets \left|\nu_{i-1}-\nu_{i}\right|$.
		\If{$\Delta_2 \geq \Delta_1$}
		\If{$i-2 = s$}
		\State Store $\left|\nu_{i-2}-\nu_{i-1}\right|$ as a half cycle:
		\State $i_c \gets i_c + 1 $,  $SoC_{av,i_c}^{cyc} \gets (\nu_{i-2}+\nu_{i-1})/2$, $DoD_{i_c} \gets \Delta_1$,  $Q_{i_c} \gets Q_{i_c-1} + \Delta_1 C / 2$.\label{alg:rain_halfcyc}
		\State Remove $\nu_{i-2}$ from $\bm{\nu}$, re-index and set $s \gets s+1, i \gets i-1$.
		\Else
		\State Store $\left|\nu_{i-2}-\nu_{i-1}\right|$ as a full cycle:
		\State $i_c \gets i_c + 1 $,  $SoC_{av,i_c}^{cyc} \gets (\nu_{i-2}+\nu_{i-1})/2$, $DoD_{i_c} \gets \Delta_1$,  $Q_{i_c} \gets Q_{i_c-1} + \Delta_1 C$.
		\State Remove $\nu_{i-2}$ and $\nu_{i-1}$ from $\bm{\nu}$, re-index and set $i \gets i-2$.
		\EndIf
		\Else
		\State $i \gets i + 1$.
		\EndIf
		\EndWhile
		\State Store all remaining differences $\left|\nu_{i-1}-\nu_{i}\right|$ in $\bm{\nu}$ as a half cycle, according to step \ref{alg:rain_halfcyc}.
	\end{algorithmic}
\end{algorithm}

\section*{References}
\bibliography{bibfile_APEN}

\end{document}